\documentclass[aps,prb,twocolumn,floatfix]{revtex4}

\usepackage{graphicx}
\usepackage{amssymb,amsfonts,amsmath}
\usepackage{epsf}
\usepackage{subfigure}
\usepackage{epstopdf}
\usepackage{mathrsfs}

\font\helvb=cmssbx12

\newcommand{\be}{\begin{equation}}
\newcommand{\ee}{\end{equation}}
\newcommand{\bea}{\begin{eqnarray}}
\newcommand{\eea}{\end{eqnarray}}

\begin{document}

\title{\bf Effective fluctuation theorems for electron transport in a double quantum dot coupled to a quantum point contact}

\author{Gregory \surname{Bulnes Cuetara}$^1$}
\email{gbulnesc@ulb.ac.be}
\author{Massimiliano Esposito$^2$}
\email{massimiliano.esposito@uni.lu}
\author{Gernot Schaller$^3$}
\email{gernot.schaller@tu-berlin.de}
\author{Pierre Gaspard$^1$}
\email{gaspard@ulb.ac.be}

\affiliation{$^1$ Center for Nonlinear Phenomena and Complex Systems, Universit\'e Libre de Bruxelles, Code Postal 231, Campus Plaine, B-1050 Brussels, Belgium\\
$^2$ Complex Systems and Statistical Mechanics, University of Luxembourg, L-1511 Luxembourg, Luxembourg\\
$^3$ Institut f\"ur Theoretische Physik, Technische Universit\"at Berlin, Hardenbergstra{\ss}e 36, D-10623 Berlin, Germany}

\begin{abstract}
A theoretical study is reported of electron transport at finite temperature in a double quantum dot (DQD)
capacitively coupled to a quantum point contact (QPC).  Starting from a Hamiltonian model, a master equation is obtained for the stochastic process taking place in the DQD while the QPC is at or away from equilibrium, allowing us to study the backaction of the QPC onto the DQD.  The QPC is treated non-perturbatively in our analysis.  Effective fluctuation theorems are established for the full counting statistics of the DQD current under different limiting conditions.  These fluctuation theorems hold with respect to an effective affinity characterizing the nonequilibrium environment of the DQD and differing from the applied voltage if the QPC is out of equilibrium.  The effective affinity may even change its sign if the Coulomb drag of the QPC reverses the DQD current.  The thermodynamic implications of the effective fluctuation theorems are discussed.
\end{abstract}

\noindent 

\vskip 0.5 cm

\maketitle

\section{Introduction}

Recent studies of electron transport in quantum mesoscopic devices, both experimental\cite{UGMSFS10,KRBMGUIE12,SYTMAP12} and theoretical,\cite{TN05,EHM07,HEM07,SU08,FB08,SLSB10,GUMS11,BEG11,KSB11} have revealed symmetries in the full counting statistics (FCS) of charge transfers that are the consequence of the so-called fluctuation theorem (FT).\cite{ECM93,GC95,K98,LS99,AG06,AG07JSP,AGMT09,EHM09,CTH11}   According to this theorem, the probabilities of forward and backward transfers have a ratio going exponentially with the numbers of transfers and the differences of electric potentials driving the currents.  This result finds its origin in the microreversibility of the underlying quantum dynamics and constitutes the basis for understanding the nonequilibrium thermodynamic properties of quantum transport at the mesoscopic level.

Typical experiments on FCS are carried out with quantum dots (QDs) capacitively coupled to an auxiliary circuit playing the role of charge detector and often taken as a quantum point contact (QPC).\cite{GLSSISEDG06,GLSSIEDG09}  Due to the electrostatic Coulomb interaction, the current in the QPC is sensitive to the electronic occupancy of the QDs, thus allowing the measurement of single-electron transitions. Moreover, by coupling the QPC asymmetrically to the QDs, it is possible to infer the directionality of the flow of charges across the QD system, providing the FCS of single-electron transfers.\cite{FHTH06,GSLIEDG07,UFFHH12}

With these devices, several experiments have established that the FCS obeys the symmetry predicted by the fluctuation theorem.\cite{UGMSFS10,KRBMGUIE12}  Fundamentally, the FT is bivariate and holds for the two currents in the QD and detector circuits and it is remarkable that the symmetry of the FT is observed for the sole current in the QD circuit.  However, the backaction of the detector onto the QD circuit modifies the symmetry by shifting the value of the voltage across the QD circuit to an effective value.  Since this effective value is experimentally accessible, a key issue is to understand how this value depends on the capacitive coupling between the detector and the QD circuit, as well as on the nonequilibrium driving forces.

The purpose of the present paper is to address this issue in the case of a double quantum dot (DQD) weakly coupled to two electrodes and probed by a QPC detector sensitive to the electronic occupation of the DQD via Coulomb interaction.  The currents are driven by the two electric potential differences applied to both conduction channels.  We use a non-perturbative analysis for the QPC circuit that is considered in fully nonequilibrium regimes.

First, we show that, at finite and homogeneous temperature, the QPC behaves as a source of Bose-like fluctuations driving transitions between the charge eigenstates of the DQD.  Our result is consistent, in the low-temperature limit, with the experimental observation of a threshold for the current induced in the DQD channel as a function of the bias across the QPC.\cite{GSLIEDG07}  This effect is directly related to the Coulomb drag exerted by the QPC onto the DQD current.\cite{SLSB10,LK08,MT09}

Secondly, we demonstrate the emergence of effective fluctuation theorems for the current in the sole DQD under different experimentally relevant conditions.  These conditions suppose that the QPC is faster than the DQD.  An effective FT holds if the tunneling rate between the two quantum dots composing the DQD is smaller than their tunneling rates with the electrodes.  Another effective FT is obtained if the QPC induces transitions between the DQD internal states at a rate faster than the DQD charging and discharging rates.  In every case, we investigate how the effective FT can characterize the DQD and its capacitive coupling to the QPC.

The paper is organized as follows.  In Section \ref{Model}, we present the model of a DQD coupled to a QPC, its Hamiltonian description, and the master equation ruling the stochastic process of electron transport.  In Section \ref{FCS}, the FCS of the DQD current is obtained in terms of the cumulant generating function. The average current in the DQD is investigated with and without a bias in the QPC, showing that the QPC may induce a strong backaction effect onto the DQD current and that the Coulomb drag of the QPC may even reverse the DQD current.  An effective affinity is introduced to characterize the nonequilibrium driving forces acting on the DQD. In Section \ref{FT}, the symmetry of the FT is shown to hold effectively under different conditions of experimental relevance and thermodynamic implications are presented.  Section \ref{Summary} contains the summary of the results and the conclusions.

\section{Double quantum dot coupled to a quantum point contact}
\label{Model}

\begin{figure}[htbp]
\centerline{\includegraphics[width=7cm]{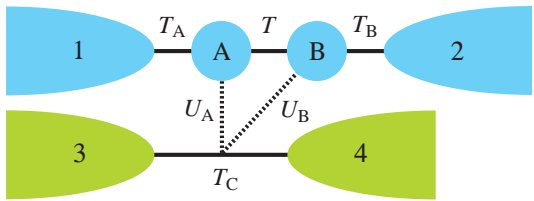}}
\caption{Schematic representation of a DQD capacitively coupled to a QPC.  The DQD is composed of the two quantum dots A and B that are coupled together as well as to the electrodes 1 and 2.  The QPC is coupling the electrodes 3 and 4.  The solid lines depict the couplings by tunneling and the dashed lines the capacitive couplings between the QPC and each QD.  The symbols are explained in the text.}
\label{fig1}
\end{figure}

A theoretical description for a DQD capacitively coupled to a QPC is set up starting from the Hamiltonian of the system, which is schematically depicted in Fig.~\ref{fig1}.\cite{GUMS11,OLY10}  Without the capacitive coupling to the DQD, the Hamiltonian model of the QPC is solved non perturbatively, leading to the Landauer-B\"uttiker formula for its average current and allowing us to calculate the correlation functions of its properties when it is in an arbitrary nonequilibrium steady state.  On the other hand, the capacitive coupling of the DQD to the QPC, as well as the coupling of the DQD to its reservoirs by direct tunneling are treated perturbatively at second order in the corresponding coupling parameters and with the rotating-wave approximation.  In this way, a master equation is obtained for the transitions between the internal states of the DQD.  The state of double occupancy is supposed to lie high enough in energy to play a negligible role.  In order to obtain the transition rates as explicitly as possible in terms of the parameters of the Hamiltonian operator, we consider tight-binding models for the reservoirs and the QPC.\cite{T01,S07}  Moreover, the wide-band approximation is used for the QPC.  As in our previous work,\cite{BEG11} the capacitances of the tunneling junctions between the QDs and the reservoirs are absent since our interest is here focused on the nonequilibrium conditions influencing the transport process.

\subsection{The Hamiltonian}

The DQD is supposed to be composed of two quantum dots in series that can exchange electrons by direct tunneling.  This system is modeled in the local basis by the Hamiltonian 
\be
H_{\rm AB} = \epsilon_{\rm A} \, d_{\rm A}^{\dagger} d_{\rm A} + \epsilon_{\rm B} \, d_{\rm B}^{\dagger} d_{\rm B} + T \left( d_{\rm A}^{\dagger} d_{\rm B} +  d_{\rm B}^{\dagger} d_{\rm A}\right)
\label{H_DQD}
\ee
where $d_{\rm A}$ and $d_{\rm B}$ are the annihilation operators of an electron in the corresponding dot.  The same Hamiltonian holds for both spin orientations, which are treated similarly and thus implicitly in our notations.  The energy of one electron in the dot A (resp. the dot~B) is equal to $\epsilon_{\rm A}$ (resp.~$\epsilon_{\rm B}$).  The tunneling amplitude between both dots is denoted $T$.  The diagonalization of this Hamiltonian is carried out in Appendix~\ref{AppA}.

The DQD is connected to the reservoirs $j=1,2$ and the QPC to the reservoirs $j=3,4$.
The reservoirs can be modeled by tight-binding Hamiltonians such as
\be
H_j = -\gamma \sum_{l=0}^{\infty} \left(d_{j,l}^{\dagger} d_{j,l+1} +  d_{j,l+1}^{\dagger} d_{j,l}\right)
\label{H_j}
\ee
where $d_{j,l}$ denotes the annihilation operator for an electron on the site of index $l\in{\mathbb N}$ in the $j^{\rm th}$ reservoir.  The advantage of such models is that these Hamiltonian operators
are exactly diagonalizable as shown in Appendix~\ref{AppB}.  The parameter $\gamma>0$ determines the width of the allowed energy band according to the dispersion relation $\epsilon_k=-2\gamma \cos k$ with the wavenumber $0\leq k \leq \pi$.  The band width is thus equal to $\Delta\epsilon=4\gamma$.
For simplicity, the parameter $\gamma$ is supposed to be common to every reservoir.

The DQD is coupled by tunneling to the reservoirs $j=1,2$ with the following interaction operators:
\bea
&& V_{1{\rm A}} = T_{\rm A} \left(d_{1,0}^{\dagger} d_{\rm A} +  d_{\rm A}^{\dagger} d_{1,0}\right) \label{V1A}\\
&& V_{2{\rm B}} = T_{\rm B} \left(d_{2,0}^{\dagger} d_{\rm B} +  d_{\rm B}^{\dagger} d_{2,0}\right) \label{V2B}
\eea
where $T_{\rm A}$ denotes the tunneling amplitude between the dot A and the reservoir $j=1$, while $T_{\rm B}$ is the tunneling amplitude between the dot B and the reservoir $j=2$.

The Hamiltonian of the QPC is taken as
\be
H_{\rm C} = H_3 + H_4 + T_{\rm C} \left(d_{3,0}^{\dagger} d_{4,0} +  d_{4,0}^{\dagger} d_{3,0}\right)
\label{H_C}
\ee
Since this Hamiltonian is quadratic in the annihilation-creation operators, it is exactly diagonalizable as shown in Appendix~\ref{AppC}, which provides the electronic scattering properties of the QPC.

The capacitive coupling between the DQD and the QPC is described by the interaction
\be
V_{\rm ABC} = \left(U_{\rm A} \, d_{\rm A}^{\dagger} d_{\rm A} + U_{\rm B} \, d_{\rm B}^{\dagger} d_{\rm B}\right)\left(d_{3,0}^{\dagger} d_{4,0} +  d_{4,0}^{\dagger} d_{3,0}\right)
\label{VABC}
\ee
where $U_{\rm A}$ and $U_{\rm B}$ are the parameters characterizing the electrostatic Coulomb interaction between the electrons in the quantum dots and the ones at the edges of the reservoirs $j=3,4$.\cite{OLY10}

Finally, the total Hamiltonian reads
\be
H = H_1 + V_{1{\rm A}} + H_{\rm AB} + V_{2{\rm B}} + H_2 + H_{\rm C} + V_{\rm ABC}
\label{H_tot}
\ee
which can be rewritten as
\be
H= H_{\rm S} + H_{\rm R} + V
\ee
where
\be
H_{\rm S} =H_{\rm AB}
\ee
is the Hamiltonian of the subsystem formed by the DQD,
\be
H_{\rm R} = H_1 + H_2 + H_{\rm C}
\ee
is the Hamiltonian of the environment of the subsystem 
including the reservoirs $j=1,2$ in contact with the DQD and the QPC,
and
\be
V = V_{1{\rm A}} + V_{2{\rm B}} + V_{\rm ABC}
\label{V}
\ee
is the interaction between both parts, which is treated perturbatively at second order in the coupling parameters $T_{\rm A}$, $T_{\rm B}$, $U_{\rm A}$, and $ U_{\rm B}$.

\subsection{The Hamiltonian in the DQD eigenbasis}

The Hamiltonian of the DQD can be diagonalized in the basis of its eigenstates $\{\vert s\rangle\}$ as
\be
H_{\rm S}=\sum_s \epsilon_s \, \vert s \rangle  \langle s \vert
\ee
In the following, we assume for simplicity that the only states entering the dynamics are the empty state $\vert 0 \rangle$ and the single-charge eigenstates $\vert + \rangle$ and $\vert - \rangle$. The expressions of these eigenstates and the corresponding energy eigenvalues are given in Appendix~\ref{AppA}.  

In the eigenbasis, the interaction operators (\ref{V1A}) and (\ref{V2B}) can be expressed as
\bea
&& V_{1{\rm A}} = \sum_{s=\pm} T_{1s} \left( \vert s \rangle \langle 0 \vert \, d_{1,0} + d_{1,0}^{\dagger} \, \vert 0 \rangle \langle s \vert \right) \label{V1A_bis}\\
&& V_{2{\rm B}} = \sum_{s=\pm} T_{2s} \left( \vert s \rangle \langle 0 \vert \, d_{2,0} + d_{2,0}^{\dagger} \, \vert 0 \rangle \langle s \vert \right) \label{V2B_bis}
\eea
where $T_{js}$ are the tunneling amplitudes in the eigenbasis given in Appendix~\ref{AppA} in terms of the parameters of the local basis.

The interaction with the QPC is similarly expressed as
\be
V_{\rm ABC} = \left(\sum_{s,s'=\pm} U_{ss'} \, \vert s \rangle \langle s' \vert\right)\left(d_{3,0}^{\dagger} d_{4,0} +  d_{4,0}^{\dagger} d_{3,0}\right)
\label{VABC_bis}
\ee
showing that the QPC can induce transitions between the equal-charge eigenstates $\vert + \rangle$ and $\vert - \rangle$ of the DQD. The coupling coefficients $U_{ss'}$ characterize the strength of the Coulomb interaction between the DQD and the QPC and are also expressed in Appendix~\ref{AppA} with the parameters of the local basis.

An important point to note is that the coupling parameters in the capacitive interaction (\ref{VABC_bis}) satisfy
\be
U_{+-}=U_{-+}
\ee
so that the transitions induced by the QPC between the eigenstates $\vert + \rangle$ and $\vert - \rangle$ have equal amplitudes.  There is thus no direction favored in the Hamiltonian.  Moreover, these coupling parameters are proportional to $U_{\rm A}-U_{\rm B}$, which characterizes the degree of asymmetry in the capacitive coupling between the QPC and the DQD (see Fig.~\ref{fig1}).  A direct consequence of this fact is the vanishing of the backaction if the QPC is symmetrically coupled to the DQD.

\subsection{The currents}

The total Hamiltonian (\ref{H_tot}) commutes separately with the total numbers of electrons in each conduction channel:
\bea
&& \left[ H, N_1+N_2+N_{\rm AB}\right] = 0 \\
&& \left[ H, N_3+N_4\right] = 0 
\eea
where
\be
N_j = \sum_{l=0}^{\infty} d_{j,l}^{\dagger} d_{j,l} 
\label{N_j}
\ee
is the number of electrons in the $j^{\rm th}$ reservoir and
\be
N_{\rm AB} = d_{\rm A}^{\dagger} d_{\rm A} + d_{\rm B}^{\dagger} d_{\rm B}
\ee
the number of electrons in the DQD.  Therefore, the electron current is conserved separately in the DQD circuit, as well as in the QPC circuit.

The current in the DQD circuit can be defined as the rate of decrease of the electron number in the reservoir $j=1$ by
\be
J_{\rm D} \equiv -\frac{dN_1}{dt} = i\left[ N_1,H\right] = i \, T_{\rm A} \left(d_{1,0}^{\dagger} d_{\rm A}-d_{\rm A}^{\dagger} d_{1,0}\right)
\ee
in units where Planck's constant is equal to $\hbar=1$.
The current in the QPC is similarly defined as the rate of decrease of the electron number in the reservoir $j=3$ by
\be
J_{\rm C} \equiv -\frac{dN_3}{dt} = i\left[ N_3,H\right] 
\ee
The average values of the electric currents are thus given by
\be
I_{\rm D} = e \langle J_{\rm D}\rangle \qquad\mbox{and}\qquad I_{\rm C} = e \langle J_{\rm C}\rangle
\ee
where $e$ is the electron charge.

The reservoirs are assumed to be initially in grand-canonical statistical ensembles at homogeneous temperature so that the inverse temperature $\beta$ is uniform across the whole system.  However, the reservoirs have different chemical potentials given by $\mu_j$ with $j=1,2,3,4$.  Because of the separate charge conservation in both circuits, the nonequilibrium conditions of this isothermal system are specified by two dimensionless affinities
\bea
&& A_{\rm D} = \beta(\mu_1-\mu_2) = \beta \Delta\mu_{\rm D}= \beta e V_{\rm D} \label{A_D}\\
&& A_{\rm C} = \beta(\mu_3-\mu_4) = \beta \Delta\mu_{\rm C}= \beta e V_{\rm C} \label{A_C}
\eea
corresponding to the differences of electric potentials in the two circuits, respectively $V_{\rm D}=\Delta\mu_{\rm D}/e$ and $V_{\rm C}=\Delta\mu_{\rm C}/e$.

\subsection{The master equation}

The interaction (\ref{V}) between the DQD and the rest of the system will be assumed to be weak and treated perturbatively to obtain the Markovian master equation ruling the time evolution of the probabilities $\{p_s(n,t)\}$ to find the DQD in one of its three eigenstates $\{\vert s \rangle\}_{s=0,\pm}$ while $n$ electrons have been transferred from the reservoir $j=1$.

These probabilities are given by
\be
p_s(n,t) = \sum_{n_1} {\rm tr} \left[ \rho_{n_1}(t) \vert s \rangle \langle s \vert \, \delta_{N_1,n_1-n}\right]\label{p_s}
\ee
where $\rho_{n_1}(t)$ is the density operator of the total system provided that the reservoir $j=1$ contains $N_1=n_1$ electrons at the initial time $t=0$ and normalized according to
\be
\sum_{n_1} {\rm tr}\, \rho_{n_1}(t)=1
\ee  
The time evolution of the density operator $\rho=\rho_{n_1}$ is ruled by the Landau-von~Neumann equation
 \be
 i \,\partial_t \, \rho = [ H,\rho] = [H_0,\rho] + [ V,\rho]
 \ee
 where $H_0=H_{\rm S}+H_{\rm R}$ is the Hamiltonian operator of the uncoupled system with $H_{\rm S}=H_{\rm AB}$ and $H_{\rm R}=H_1+H_2+H_{\rm C}^{(+)}+H_{\rm C}^{(-)}$ while $V$ is the interaction operator (\ref{V}).  The operators $H_{\rm C}^{(\pm)}$ are obtained from the diagonalization of the QPC Hamiltonian (\ref{H_C}) in Appendix~\ref{AppC}.  
 $H_{\rm C}^{(+)}$ denotes the Hamiltonian (\ref{HC+}) for the waves in the QPC with positive wavenumbers $q>0$ coming from the reservoir $j=3$ and $H_{\rm C}^{(-)}$ the Hamiltonian (\ref{HC-}) for those with negative wavenumbers $q<0$ coming from the reservoir $j=4$.  
 
 The initial density operator is given by
\be
\rho_{n_1}(0) = \rho_{\rm S}(0)\; \rho_{\rm R}(0) \; \delta_{N_1,n_1}
\ee
with an arbitrary statistical mixture $\rho_{\rm S}(0)$ for the subsystem and
 \bea
 &&\rho_{\rm R}(0)= \frac{1}{\Xi_1} \, {\rm e}^{-\beta(H_1-\mu_1N_1)} \frac{1}{\Xi_2} \, {\rm e}^{-\beta(H_2-\mu_2N_2)} \nonumber\\
 && \times \frac{1}{\Xi_{\rm C}^{(+)}} \, {\rm e}^{-\beta(H_{\rm C}^{(+)}-\mu_3N_{\rm C}^{(+)})}
\; \frac{1}{\Xi_{\rm C}^{(-)}} \, {\rm e}^{-\beta(H_{\rm C}^{(-)}-\mu_4N_{\rm C}^{(-)})}\qquad
\label{rho_R0}
 \eea
 for the environment.\cite{T01,LS11}  This initial density operator expresses the assumption that the reservoirs $j=1$ and $j=2$ in contact with the DQD are at different chemical potentials $\mu_1$ and $\mu_2$ while the QPC is in a steady state, possibly out of equilibrium, with electrons flowing in one direction from the reservoir $j=3$ at the chemical potential $\mu_3$ and in the opposite direction from the reservoir $j=4$ at the chemical potential $\mu_4$, the whole system being at the uniform inverse temperature~$\beta$.
 
Using perturbation theory at second order together with the rotating-wave approximation, it is well known\cite{CDG96,BP02,GZ04,SKB10} that, in the Markovian limit, the master equation for the probabilities (\ref{p_s}) reads
\be
\frac{d}{dt}p_s(n,t) = \sum_{\nu}\sum_{s'} \left[L_{ss'}^{(\nu)} \hat E^{\nu} \, p_{s'}(n,t) - L_{s's}^{(\nu)} \, p_s(n,t)\right]
\label{master}
\ee
where $\nu=+1,0,-1$ is the number of electrons transferred to the reservoir $j=1$ during the transition $s'\to s$.  The operators
\be
\hat E^{\nu} \equiv \exp\left(\nu \frac{\partial}{\partial n}\right)
\label{E}
\ee
change the number $n$ of transferred electrons by the quantity $\nu$ according to $\hat E^{\nu}\phi(n)=\phi(n+\nu)$ for any function $\phi(n)$.  The Markovian master equation (\ref{master}) describes processes taking place over time scales $\Delta t$ that are longer than the correlation time of the environment: $\Delta t \gg \tau^{\rm (C)}$.  This correlation time is estimated as the characteristic time scale over which the environmental correlation functions (\ref{C_ab}) have decayed to half their value.  The correlation time is thus inversely proportional to the band width of the reservoirs: $\tau^{\rm (C)} \sim 2\pi\hbar/(4\gamma)$.  

If the probabilities $\{p_s(n,t)\}_{s=0,\pm}$ are gathered in the array
\be
{\bf p}(n,t)=
\left(
\begin{array}{c}
p_{0}(n,t) \\
p_{+}(n,t) \\
p_{-}(n,t) 
\end{array}
\right)
\ee
the master equation (\ref{master}) can be written as
\begin{widetext}
\be
\partial_t\,{\bf p}(n,t) = \hat{\mbox{\helvb L}} \cdot {\bf p}(n,t) 
\label{master-m}
\ee
with the matrix
\be
\hat{\mbox{\helvb L}}
=
\left(
\begin{array}{ccc}
-a_{1+}-a_{2+}-a_{1-}-a_{2-} & b_{1+}\,\hat{E}^+ +b_{2+} & b_{1-}\,\hat{E}^+ +b_{2-} \\
a_{1+}\,\hat{E}^- +a_{2+} & -b_{1+}-b_{2+}-c_{-+} & c_{+-} \\
a_{1-}\,\hat{E}^- +a_{2-} & c_{-+} & -b_{1-}-b_{2-}-c_{+-} 
\end{array}
\right)
\label{L}
\ee
\end{widetext}
which defines an operator since $\hat E^{\pm}p_s(n,t)=p_s(n\pm 1,t)$.

The perturbative calculation of the transition rates is carried out in Appendices~\ref{AppD} and \ref{AppE} within the wide-band approximation.  They include the charging and discharging transition rates between the reservoirs $j=1,2$ and the eigenstates $s=\pm$ given by
\bea
&& a_{js} = \Gamma_{js} \, f_j(\epsilon_s) \label{ajs} \\
&& b_{js} = \Gamma_{js} \, \left[1-f_j(\epsilon_s)\right] \label{bjs}
\eea in terms of the Fermi-Dirac distribution
\be
f_j(\epsilon) = \frac{1}{{\rm e}^{\beta(\epsilon-\mu_j)}+1}
\label{FD}
\ee
and the rates
\be
\Gamma_{js} = 4\pi g \, T_{js}^2
\label{Gjs_wb}
\ee
where $T_{js}$ are the tunneling amplitudes (\ref{T1+})-(\ref{T2-}) and $g$ is the local density of states (\ref{LDOS}).  We notice that the charging and discharging rates obey the local detailed balance conditions:
\be
\frac{a_{js}}{b_{js}}={\rm e}^{-\beta(\epsilon_s-\mu_j)}
\ee
The thermal energy is assumed to be larger than the natural width of the DQD energy levels, $\beta\hbar(\Gamma_{1s}+\Gamma_{2s})\ll 1$, in consistency with the neglect of resonance effects by second-order perturbation theory.\cite{B91}

The coefficients of the matrix (\ref{L}) also include the rates of the transitions induced by the capacitive coupling to the QPC:
\be
c_{ss'} =\frac{8\pi g^2 U_{ss'}^2}{(1+\kappa)^2} \ \left[ \frac{\omega_{ss'}-\Delta\mu_{\rm C}}{{\rm e}^{\beta(\omega_{ss'}-\Delta\mu_{\rm C})}-1} + \frac{\omega_{ss'}+\Delta\mu_{\rm C}}{{\rm e}^{\beta(\omega_{ss'}+\Delta\mu_{\rm C})}-1}\right]
\label{c+-}
\ee
where $s=-s'=\pm$, $\omega_{ss'}=\epsilon_s-\epsilon_{s'}$, $\Delta\mu_{\rm C}=\mu_3-\mu_4$, $g$ is the common local density of states (\ref{LDOS}), and $\kappa$ is the dimensionless contact transparency (\ref{kappa}) of the QPC.\cite{GK06}  These transition rates are proportional to the intensity of the capacitive coupling: $U_{+-}^2=U_{-+}^2=(U_{\rm A}-U_{\rm B})^2\sin^2\theta/4$.  The expression (\ref{c+-}) shows the Bose-like character of the random transitions due to the backaction of the QPC onto the DQD circuit.  If the QPC is at equilibrium with $\Delta\mu_{\rm C}=0$, these transition rates satisfy the condition of local detailed balance:
\be
\frac{c_{+-}}{c_{-+}}={\rm e}^{-\beta\omega_{+-}} \qquad\mbox{for} \qquad \Delta\mu_{\rm C}=0
\label{c_ratio_eq}
\ee
However, this condition is not satisfied under general nonequilibrium conditions $\Delta\mu_{\rm C}\neq 0$ for the QPC.

We notice that, if the QPC is at a uniform temperature different from the DQD temperature, the inverse temperature $\beta$ in the rates~(\ref{c+-}) should be replaced by the inverse temperature $\beta_{\rm C}$ of the QPC and the master equation (\ref{master-m})-(\ref{L}) would again apply.

\subsection{The fluctuating current in the QPC}

The master equation (\ref{master}) rules the stochastic process of electrons jumping between the DQD and the reservoirs.  This stochastic process can be simulated by Monte Carlo algorithms to obtain random histories $\{\vert s_t\rangle,n_t\}_{t\in{\mathbb R}}$ for the system.\cite{GZ04,GMWS01}  Due to the capacitive coupling, the occupancy of the quantum dots by electrons modulates the current in the QPC.  Indeed, if the DQD is in the instantaneous state $\vert s_t\rangle$, the capacitive interaction (\ref{VABC}) modifies the tunneling amplitude $T_{\rm C}$ of the QPC into the time-dependent effective amplitude
\be
\tilde T_{\rm C}(t) = T_{\rm C} + \langle s_t\vert U_{\rm A} \, d_{\rm A}^{\dagger} d_{\rm A} + U_{\rm B} \, d_{\rm B}^{\dagger} d_{\rm B}\vert s_t\rangle
\label{T_C(t)}
\ee

If the correlation time of the QPC is shorter than the dwell time of the electrons in the DQD, the current in the QPC can be estimated with the Landauer-B\"uttiker formula with a transmission probability given in terms of the time-dependent tunneling amplitude (\ref{T_C(t)}).
Over time scales longer than the QPC correlation time, $\Delta t\gg \tau^{\rm (C)}$, the current in the QPC is thus given for each spin orientation by
\be
\langle J_{\rm C} \rangle_t = \frac{1}{2\pi} \int_{-2\gamma}^{+2\gamma} d\epsilon \; \tilde{\cal T}_{\epsilon}(t) \left[ f_3(\epsilon)-f_4(\epsilon)\right]
\ee
in terms of the transmission probability~(\ref{T_E}) but with the tunneling amplitude $T_{\rm C}$ replaced by the time-dependent expression (\ref{T_C(t)}).  

If the DQD is occupied, Coulomb repulsion raises the barrier in the QPC, thus, lowering its current.  If the DQD is temporarily in the state $\vert +\rangle$, the tunneling amplitude takes the value $\tilde T_{\rm C}(t) = T_{\rm C} +U_{++}$ with the Coulomb repulsion (\ref{U++}).  Instead, the tunneling amplitude is equal to $\tilde T_{\rm C}(t) = T_{\rm C} +U_{--}$ with (\ref{U--}) if the DQD state is $\vert -\rangle$.  The QPC current is thus sensitive to the directionality of the electron jumps in the DQD if the Coulomb repulsion is asymmetric between the two states $\vert \pm \rangle$, which requires that the mixing angle is not close to $\theta=\pi/2$ otherwise $U_{++}=U_{--}$.

\section{Full counting statistics of electron transport in the DQD}
\label{FCS}

\subsection{Cumulant generating function}

The counting statistics of electron transfers in the DQD is fully characterized in terms of the generating function of the statistical cumulants for the random number $n$ of electrons transferred from the reservoir $j=1$:
\be
Q(\lambda) \equiv \lim_{t\to\infty} - \frac{1}{t} \ln \left\langle \exp(-\lambda n) \right\rangle_t
\label{Q}
\ee
where the average $\langle\cdot\rangle_t$ is carried out over the probability distribution $p(n,t)=\sum_s p_s(n,t)$ that $n$ electrons have been transferred during the time interval $[0,t]$.
The probabilities $p_s(n,t)$ denote the solutions of the master equation (\ref{master}) starting from initial conditions $p_s(n,0)=P_s \,\delta_{n,0}$ such that $P_s=\sum_n p_s(n,t)$ are the stationary probabilities of the DQD internal states.  It is known that the cumulant generating function is given by the leading eigenvalue of the matrix 
\be
\mbox{\helvb L}(\lambda) \equiv {\rm e}^{-\lambda n}\, \hat{\mbox{\helvb L}} \; {\rm e}^{\lambda n}
\label{L_lambda}
\ee
obtained with the substitutions $\hat E^{\pm} \to {\rm e}^{\pm\lambda}$ in the matricial operator (\ref{L}).  The generating function is thus obtained by solving the eigenvalue problem:
\be
\mbox{\helvb L}(\lambda)\cdot {\bf v} = - Q(\lambda) \, {\bf v}
\label{eigenvalue_eq}
\ee
where ${\bf v}={\bf v}(\lambda)$ is the associated eigenvector.  Since the matrix elements depend on the chemical potentials of the reservoir, the generating function characterizes the current fluctuations in a stationary state of the DQD that is generally out of equilibrium.  The complete equilibrium state is reached if both affinities (\ref{A_D}) and (\ref{A_C}) are vanishing.

The average current in the DQD as well as the higher cumulants are given by taking successive derivatives of the generating function with respect to the counting parameter $\lambda$.  In particular, the average current is obtained as
\be
\langle J_{\rm D}\rangle = \frac{\partial Q}{\partial\lambda}\Big\vert_{\lambda=0}
\label{J_D-Q}
\ee
in a stationary state of the DQD.  The second derivative gives the diffusivity of the current fluctuations around its average value.

\subsection{The average current in the DQD}

If $\{P_s\}$ denote the stationary probabilities to find the DQD in one of its three internal states $\{\vert s\rangle\}$, the average current is given by
\be
\langle J_{\rm D}\rangle = (a_{1+}+a_{1-}) \, P_0 - b_{1+}\, P_+ - b_{1-} \, P_-
\label{J_D}
\ee
Indeed, the net average current from the reservoir $j=1$ has two positive contributions due to the charging transitions $\vert 0 \rangle \to \vert\pm\rangle$ from the reservoir $j=1$ and two negative contributions due to the discharging transitions $\vert\pm\rangle \to \vert 0 \rangle$ back to the reservoir $j=1$.  The stationary probabilities $\{P_s\}$ form the eigenvector of the matrix $\mbox{\helvb L}(\lambda=0)$ associated with its zero eigenvalue.

\begin{figure}[h]
\centerline{\includegraphics[width=8cm]{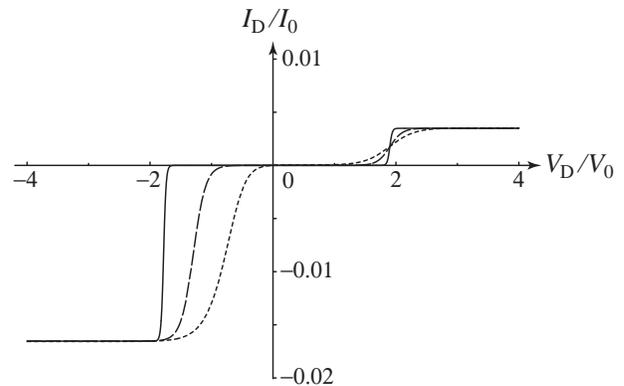}}
\caption{The average current $I_{\rm D}=e\langle J_{\rm D}\rangle$ in the DQD versus the applied potential difference, $V_{\rm D}=\Delta\mu_{\rm D}/e=(\mu_1-\mu_2)/e$, if the QPC is at equilibrium with $\Delta\mu_{\rm C} = 0$.  The inverse temperature is $\beta = 10, 20, 100$ for respectively the dotted, dashed, and continuous lines.  The other parameters are $\mu_1+\mu_2= 3.3$, $\epsilon_{\rm A} = 0.7$, $\epsilon_{\rm B}  = 1.3$, $T = 0.01$, $T_{\rm A} = T_{\rm B} = 1$, $U_{\rm A} = -2.1$, $U_{\rm B} =- 0.6$, $\kappa = 0.2$, and $g = 1$.  The units are $I_0=eT_{\rm A}^2$ and $V_0=(\epsilon_{\rm A}+\epsilon_{\rm B})/(2e)$.}
\label{fig2}
\end{figure}

Figure~\ref{fig2} shows several $I$-$V$ characteristic curves of the DQD circuit in the absence of bias in the QPC at different temperatures.  Since the QPC is at equilibrium $\Delta\mu_{\rm C}=0$, the average current $I_{\rm D}$ in the DQD vanishes with the applied potential $V_{\rm D}$.  In Fig.~\ref{fig2}, the eigenstates of the DQD have the energies $\epsilon_+\simeq 1.3$, $\epsilon_-\simeq 0.7$, and $\epsilon_0=0$.  The steps in the $I$-$V$ curves arise because every Fermi-Dirac distribution $f_j(\epsilon_s)$ undergo a similar step at the thresholds $\epsilon_s=\mu_j$ with $j=1,2$ and $s=\pm$.  {\it A priori}, thresholds are thus expected at the values $V_{\rm D}\simeq \pm 0.7$ if $\epsilon_+=\mu_j$, and $V_{\rm D}\simeq \pm 1.9$ if $\epsilon_-=\mu_j$ ($j=1,2$).  Nevertheless, only the latter ones appear in Fig.~\ref{fig2} under the condition $\Delta\mu_{\rm C}=0$.  The reason is that the capacitive coupling to the QPC favors the transitions $\vert + \rangle\to\vert - \rangle$ because of Eq.~(\ref{c_ratio_eq}) and thus depopulates the level $\vert + \rangle$.  However, the level $\vert - \rangle$ remains below the Fermi energies of both reservoirs if $0<V_{\rm D}< 1.9$ so that the current is essentially stopped in this range.

\begin{figure}[h]
\centerline{\includegraphics[width=8cm]{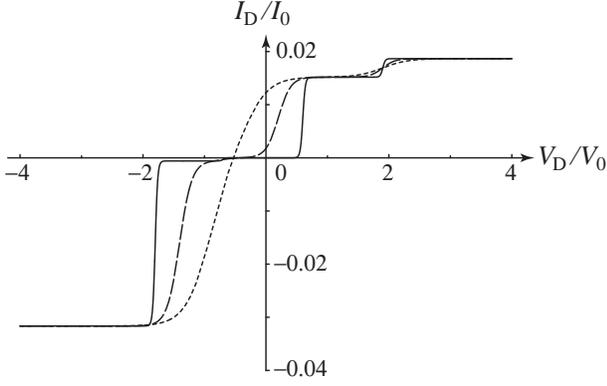}}
\caption{The average current $I_{\rm D}=e\langle J_{\rm D}\rangle$ in the DQD versus the applied potential difference, $V_{\rm D}=\Delta\mu_{\rm D}/e$, if the QPC is out of equilibrium with $\Delta\mu_{\rm C} = -2$.  The inverse temperature is $\beta = 10, 20, 100$ for respectively the dotted, dashed, and continuous lines.  The other parameters and the units are the same as in Fig.~\ref{fig2}.}
\label{fig3}
\end{figure}

In contrast, Fig.~\ref{fig3} shows that, for $\Delta\mu_{\rm C}=-2$, the current no longer vanishes with the voltage bias applied to the DQD and may even go against the bias if $V_{\rm D}\lesssim 0$.
This remarkable effect is due to the Coulomb drag that manifests itself because the QPC is out of equilibrium and capacitively coupled to the DQD circuit.  In the linear regime for small enough values of the applied voltages, the average currents are related to the affinities (\ref{A_D})-(\ref{A_C}) according to $\langle J_{m}\rangle \simeq \sum_{m'}{\cal L}_{mm'} A_{m'}$ in terms of the Onsager coefficients ${\cal L}_{mm'}$ with $m,m'={\rm C,D}$.  Because of the asymmetric capacitive coupling the coefficients ${\cal L}_{\rm CD}={\cal L}_{\rm DC}$ are non vanishing, allowing the Coulomb drag effect observed in Fig.~\ref{fig3} specially at high temperature for $\beta=10$.
If $V_{\rm D}=0$, the average current (\ref{J_D}) is proportional to
\be
\langle J_{\rm D}\rangle \propto \left(\Gamma_{1+}\Gamma_{2-}-\Gamma_{1-}\Gamma_{2+}\right)\left( c_{-+}\, {\rm e}^{\beta\epsilon_-}-c_{+-}\, {\rm e}^{\beta\epsilon_+}\right)
\label{J_D-V_D=0}
\ee
with $\Gamma_{1+}\Gamma_{2-}-\Gamma_{1-}\Gamma_{2+}=(4\pi g T_{\rm A}T_{\rm B})^2\cos\theta$.  This current vanishes if the QPC is at equilibrium when the local detailed balance condition (\ref{c_ratio_eq}) holds, which is no longer the case out of equilibrium.  We notice that the current (\ref{J_D-V_D=0}) also vanishes if the mixing angle reaches the value $\theta=\pi/2$.  The drag effect thus requires a good localization of the eigenstates in either one or the other of both dots, a condition which is met if the tunneling amplitude $T$ is not too large.  

\begin{figure}[htbp]
\centerline{\includegraphics[width=7cm]{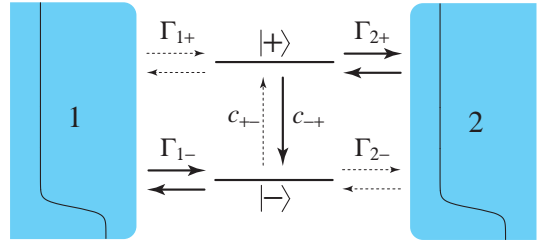}}
\caption{Schematic representation of the transitions in the DQD circuit if the QPC fulfills the condition $\vert\Delta\mu_{\rm C}\vert < \epsilon_+-\epsilon_-$ for which Eqs.~(\ref{c+-0})-(\ref{c-+0}) hold and, moreover, if $\Gamma_{1-},\Gamma_{2+}\gg\Gamma_{1+},\Gamma_{2-}$, as it is the case for $\theta\simeq\pi$.  The solid line in each reservoir depicts the corresponding Fermi-Dirac distribution.}
\label{fig4}
\end{figure}

Figure~\ref{fig3} also shows that, for $\Delta\mu_{\rm C}=-2$, the four thresholds $V_{\rm D}\simeq \pm 0.7$ and $V_{\rm D}\simeq \pm 1.9$ appear in the $I$-$V$ curves of the DQD.  This is explained by the behavior of the transition rates (\ref{c+-}) as a function of the bias in the QPC.  At low enough temperature, under the condition that
\be
\vert\Delta\mu_{\rm C}\vert < \omega_{+-}=\sqrt{(\epsilon_{\rm A}-\epsilon_{\rm B})^2+4T^2} 
\label{threshold}
\ee
the transition rates (\ref{c+-}) are given by
\bea
&& c_{+-} \simeq 0 \label{c+-0}\\
&& c_{-+} \simeq \frac{16\pi g^2 U_{ss'}^2}{(1+\kappa)^2} \, \omega_{+-} \label{c-+0}
\eea
Therefore, the rate $c_{+-}$ of the transition $\vert -\rangle\to \vert +\rangle$ vanishes if $\vert\Delta\mu_{\rm C}\vert < \epsilon_+-\epsilon_-$ so that the upper energy level $\epsilon_+$ remains depopulated, as shown in Fig.~\ref{fig4}.  

However, both transition rates $c_{\pm\mp}$ are positive if $\vert\Delta\mu_{\rm C}\vert > \epsilon_+-\epsilon_-$, allowing the upper energy level $\epsilon_+$ to become populated.  As aforementioned, the thresholds at $V_{\rm D}\simeq \pm 0.7$ correspond to the condition $\epsilon_+=\mu_j$.  Therefore, these thresholds also appear in the $I_{\rm D}$-$V_{\rm D}$ curves for $\vert\Delta\mu_{\rm C}\vert > \epsilon_+-\epsilon_-\simeq 0.6$.  This is the case in Fig.~\ref{fig3} where $\Delta\mu_{\rm C}=-2$ so that the four thresholds $V_{\rm D}\simeq \pm 0.7$ and $V_{\rm D}\simeq \pm 1.9$ are visible under such conditions.

\begin{figure}[htbp]
\centerline{\includegraphics[width=7cm]{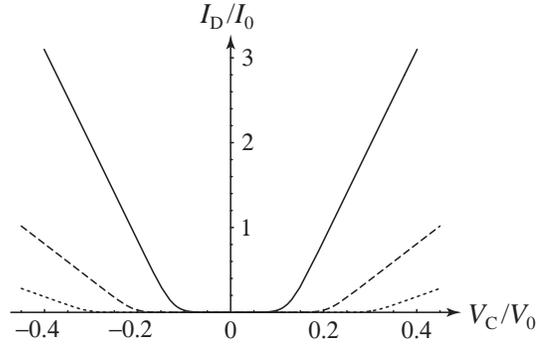}}
\caption{The average current $I_{\rm D}=e\langle J_{\rm D}\rangle$ in the DQD without applied potential, $\mu_1=\mu_2=0.11$,  versus the potential difference in the QPC, $V_{\rm C}=\Delta\mu_{\rm C}/e=(\mu_3-\mu_4)/e$.  The energy of the quantum dot B is $\epsilon_{\rm B}  = 0.21, 0.3, 0.4$ for respectively the continuous, dashed, and dotted lines.  The other parameters are $\beta = 100$, $\epsilon_{\rm A} = 0.1$, $T = 0.03$, $T_{\rm A} = T_{\rm B} = 10$, $U_{\rm A} =- 5$, $U_{\rm B} =- 1.2$, $\kappa = 0.17$, and $g = 1$.  The values of the parameters $\beta$, $\epsilon_{\rm B}-\epsilon_{\rm A}$, $T$, and $\kappa$ are estimated from experimental conditions.\cite{GSLIEDG07} 
For the given values of $T_{\rm A}=T_{\rm B}$, $U_{\rm A}-U_{\rm B}$, and $\mu_1=\mu_2$, the two largest rates are $a_{1-}\simeq 1.0$\,kHz and $b_{2+}\simeq 1.2$\,kHz in accordance with experimental data.\cite{GSLIEDG07}  The units are $I_0=e/{\rm s}$ and $V_0=1$~mV if energies are counted in meV.}
\label{fig5}
\end{figure}

Figure~\ref{fig5} shows the current in the unbiased DQD circuit versus the potential difference in the QPC for different values of the detuning $\epsilon_{\rm B}-\epsilon_{\rm A}$ between the energy levels of the quantum dots.  The DQD current remains vanishing as long as the QPC potential difference satisfies the condition~(\ref{threshold}).  Again, under this condition, the upper energy level $\epsilon_+$ is not populated because the transition $\vert -\rangle\to \vert +\rangle$ does not occur according to Eq.~(\ref{c+-0}).  Since the lower level is charged from the left-hand reservoir $j=1$ and the upper level is discharged to the right-hand reservoir $j=2$, the DQD current remains switched off.  Instead, beyond the threshold for $\vert\Delta\mu_{\rm C}\vert >\epsilon_+ -\epsilon_-=\omega_{+-}$, the upper energy level $\epsilon_+$ becomes populated thanks to the transitions induced by the QPC, which can thus exert Coulomb drag on the DQD circuit.  The DQD current is thus mainly determined by the rate $c_{+-}$.  In Fig.~\ref{fig5}, the parameter values are taken to compare with the experimental observations reported in Ref.~\onlinecite{GSLIEDG07}.  We see the remarkable agreement with Fig.~4b of this reference.

These results show that the backaction of the QPC may strongly affect the transport properties of the DQD circuit.  By enabling transitions between the states $\vert+\rangle$ and $\vert-\rangle$, the current across the DQD circuit is enhanced if the QPC is driven out of equilibrium.  The capacitive coupling between the QPC and the DQD should be asymmetric to allow the Coulomb drag to manifest itself.  Furthermore, we notice that the backaction tends to decrease if the dimensionless contact transparency $\kappa$ of the QPC increases.  A perturbative treatment of conductance in the QPC would thus overestimate the effects of backaction.

\subsection{The effective affinity}

\begin{figure}[b]
\centerline{\includegraphics[width=7cm]{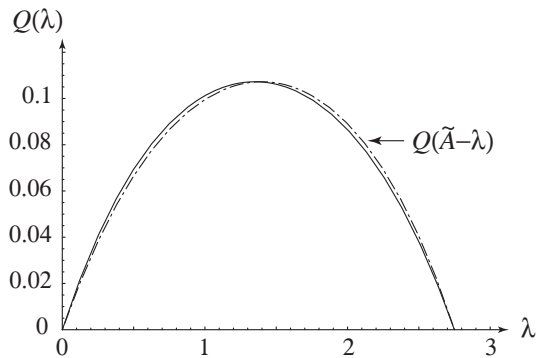}}
\caption{The cumulant generating function (\ref{Q}) versus the counting parameter $\lambda$ compared to the function transformed by the reflection $\lambda\to\tilde A-\lambda$ with the effective affinity $\tilde A=2.7495$ (dotted-dashed line) for the parameter values $\mu_1=1.5$, $\mu_2=1$, $\Delta\mu_{\rm C}=5$, $\beta = 10$, $\epsilon_{\rm A} = 0.7$, $\epsilon_{\rm B}  = 1.2$, $T = 0.5$, $T_{\rm A} = T_{\rm B} = 1$, $U_{\rm A} = -0.21$, $U_{\rm B} =- 0.06$, $\kappa = 0.2$, and $g = 1$.}
\label{fig6}
\end{figure}

\begin{figure}[b]
\centerline{\includegraphics[width=7cm]{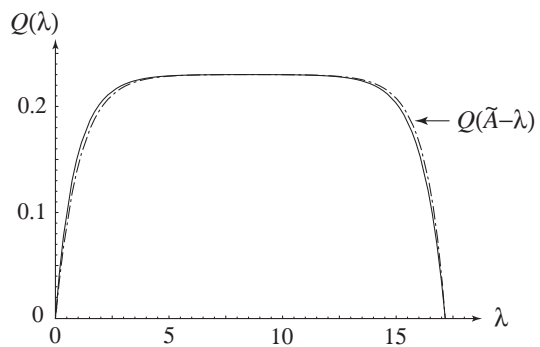}}
\caption{The cumulant generating function (\ref{Q}) versus the counting parameter $\lambda$ compared to the function transformed by the reflection $\lambda\to\tilde A-\lambda$ with the effective affinity $\tilde A=17.1636$ (dotted-dashed line) for the parameter values $\mu_1=3$, $\mu_2=1$, $\Delta\mu_{\rm C}=5$, $\beta = 10$, $\epsilon_{\rm A} = 0.7$, $\epsilon_{\rm B}  = 1.2$, $T = 0.5$, $T_{\rm A} = T_{\rm B} = 1$, $U_{\rm A} =- 0.21$, $U_{\rm B} = -0.06$, $\kappa = 0.2$, and $g = 1$.}
\label{fig7}
\end{figure}

The cumulant generating function (\ref{Q}) is shown in Figs.~\ref{fig6} and \ref{fig7} for low and high potential differences in the DQD.  As expected by its definition (\ref{Q}), the generating function vanishes at $\lambda=0$ where its slope gives the average current by Eq.~(\ref{J_D-Q}).  Moreover, the generating function also vanishes at a non-zero value of the counting parameter $\lambda=\tilde A$, which defines the {\it effective affinity}.    Since the generating function is given by the smallest root of the eigenvalue polynomial
\be
\det\left[ \mbox{\helvb L}(\lambda)+Q(\lambda)\mbox{\helvb 1}\right]=0
\ee
and $Q(\tilde A)=0$, the effective affinity can be obtained by solving
\be
\det\mbox{\helvb L}(\tilde A)= D_+\left({\rm e}^{-\tilde A}-1\right)+D_-\left({\rm e}^{\tilde A}-1\right)=0
\ee
where $D_{\pm}$ are quantities given in terms of the coefficients of the matrix (\ref{L}).
The non-trivial root is equal to $\tilde A = \ln(D_+/D_-)$, which gives the {\it effective affinity}
\begin{widetext}
\be
\tilde A =\ln \frac{a_{1+}\left[b_{2+}(b_{1-}+b_{2-}+c_{+-})+b_{2-}c_{-+}\right]+a_{1-}\left[b_{2-}(b_{1+}+b_{2+}+c_{-+})+b_{2+}c_{+-}\right]}{a_{2+}\left[b_{1+}(b_{1-}+b_{2-}+c_{+-})+b_{1-}c_{-+}\right]+a_{2-}\left[b_{1-}(b_{1+}+b_{2+}+c_{-+})+b_{1+}c_{+-}\right]}
\label{Aff_eff}
\ee
\end{widetext}


In Figs.~\ref{fig6} and \ref{fig7}, the generating function $Q(\lambda)$ is compared to $Q(\tilde A-\lambda)$.  For typical values of the parameters, the difference between $Q(\lambda)$ and $Q(\tilde A-\lambda)$ turns out to remain small in this system.  However, the symmetry under the transformation $\lambda\to\tilde A-\lambda$ is not valid in general.  Therefore, it is remarkable that there exist special conditions under which this symmetry nevertheless holds, as demonstrated in the following section.

\section{Effective fluctuation theorems}
\label{FT}

\subsection{From bivariate to univariate fluctuation theorems}

Fundamentally, a bivariate fluctuation theorem holds for both currents in the DQD and QPC circuits with respect to the basic affinities (\ref{A_D})-(\ref{A_C}).\cite{AG06,SU08,AGMT09,EHM09}  However, the conditions of observation are such that the full counting statistics of a circuit requires its coupling to another circuit so that bivariate statistics is not available.  Besides, the bivariate fluctuation theorem does not generally imply a fluctuation theorem for the sole current that is observed.  Nevertheless, conditions can be found for which univariate fluctuation theorems hold.\cite{BEG11,RJ07,GS11,MLBBS12,SSBE13}

In the present system, several such conditions exist: 

(1) If the QPC is at equilibrium, it has no other influence than an equilibrium environment so that the only source of nonequilibrium driving comes from the voltage applied to the DQD.

(2) If the tunneling amplitude between the two dots composing the DQD is small enough $\vert T\vert\ll\vert\epsilon_{\rm A}-\epsilon_{\rm B}\vert$, each dot equilibrates with its next-neighboring reservoir on a time scale that is shorter than electron transfer time scale.

(3) If the QPC induces fast transitions $\vert + \rangle \leftrightharpoons \vert - \rangle$ in the limit $c_{\pm\mp}\gg a_{js},b_{js}$, the two states $\vert +\rangle$ and $\vert -\rangle$ can be lumped together. Accordingly, the three-state process reduces by coarse graining to a two-state process, in which the DQD is either empty or singly occupied, and a fluctuation theorem always holds for two-state processes.\cite{TN05,AG06}

In these limiting cases, the difference between the cumulant generating function $Q(\lambda)$ and the transformed function $Q(\tilde A-\lambda)$ goes to zero so that the symmetry relation
\be
Q(\lambda)=Q(\tilde A-\lambda)
\label{FT-Q}
\ee
is obtained.  As a corollary, the probability $p(n,t)=\sum_s p_s(n,t)$ that $n$ electrons are transferred across the DQD during the time interval $[0,t]$ obeys the {\it fluctuation theorem}
\be
\frac{p(n,t)}{p(-n,t)}\simeq {\rm e}^{\tilde A \, n} \qquad\mbox{for}\quad t\to\infty
\label{FT1}
\ee
as proved using the theory of large deviations.\cite{T09}  The symmetry of the fluctuation theorem is established with respect to the effective affinity (\ref{Aff_eff}) taken in the limit where the equality (\ref{FT-Q}) is valid.  An important result is that the effective affinity may differ from the value (\ref{A_D}) fixed by the reservoirs alone, because of the backaction of the capacitively coupled circuit.  Here below, the effective affinity is given for the different limits where the univariate fluctuation theorem holds.

\subsection{Thermodynamic implications}

A consequence of the fluctuation theorem is that the product of the effective affinity with the average value of the current is always non negative:
\be
\tilde A \; \langle J_{\rm D}\rangle \geq 0
\label{inequal}
\ee
where $\langle J_{\rm D}\rangle=\lim_{t\to\infty}\langle n\rangle_t/t$ with $\langle n\rangle_t = \sum_{n=-\infty}^{+\infty} n \, p(n,t)$.\cite{E12}  The inequality (\ref{inequal}) constitutes a lower bound on the thermodynamic entropy production
\be
\frac{1}{k_{\rm B}}\frac{d_{\rm i}S}{dt} = A_{\rm C} \langle J_{\rm C}\rangle+A_{\rm D} \langle J_{\rm D}\rangle \geq  \tilde A \; \langle J_{\rm D}\rangle \geq 0
\label{inequal2}
\ee
as demonstrated in Appendix~\ref{AppF}.  

This lower bound reduces the thermodynamic efficiency of the energy transduction processes that the system could perform.  In particular, the Coulomb drag of the QPC may drive the DQD current against the applied voltage.  During this process, the QPC provides energy that accumulates between the reservoirs of the DQD circuit.  To characterize the balance of energy per unit time, we introduce the powers $\Pi_{m}=V_{m}I_{m}=k_{\rm B}T A_{m}\langle J_{m}\rangle$ consumed by the circuits $m={\rm C},{\rm D}$.  The power of the QPC  is positive, $\Pi_{\rm C}>0$, although the power of the DQD is negative $\Pi_{\rm D}<0$.  The thermodynamic efficiency of the process can be defined as
\be
\eta \equiv - \frac{\Pi_{\rm D}}{\Pi_{\rm C}} = -\frac{A_{\rm D} \langle J_{\rm D}\rangle}{A_{\rm C} \langle J_{\rm C}\rangle}
\label{eta}
\ee
which is positive in the regime where the Coulomb drag drives the DQD current against the applied voltage, $\eta> 0$. The general non-negativity of the thermodynamic entropy production implies the well-known upper bound $\eta\leq 1$.

Now, in the regimes where the effective fluctuation theorem (\ref{FT1}) holds, the thermodynamic entropy production has the lower bound (\ref{inequal2}) so that the efficiency is bounded as
\be
\eta \leq \frac{1}{1-\tilde A/A_{\rm D}} < 1 \qquad\mbox{if}\quad \tilde A/A_{\rm D}<0
\label{eta-bound}
\ee
The thermodynamic efficiency is thus reduced if the backaction of the QPC is strong enough to reverse the effective affinity $\tilde A$ with respect to the value $A_{\rm D}$ fixed by the voltage applied to the DQD.

The different cases where the effective fluctuation theorem (\ref{FT1}) holds are presented in the following subsections.

\subsection{The cases of univariate fluctuation theorems}

\subsubsection{The QPC is at equilibrium}

In this case, the rates of the transitions induced by the QPC obey the local detailed balance condition (\ref{c_ratio_eq}) so that the matrix (\ref{L_lambda}) obeys the symmetry relation
\be
\mbox{\helvb M}^{-1}\cdot\mbox{\helvb L}(\lambda)\cdot \mbox{\helvb M}= \mbox{\helvb L}(A_{\rm D}-\lambda)^{\rm T}
\ee
with the matrix
\be
\mbox{\helvb M} = 
\left(
\begin{array}{ccc}
1 & 0 & 0 \\
0 & {\rm e}^{-\beta(\epsilon_+-\mu_2)} & 0  \\
0 & 0 & {\rm e}^{-\beta(\epsilon_--\mu_2)}
\end{array}
\right)
\label{M}
\ee
and the affinity (\ref{A_D}).  Consequently, the leading eigenvalue giving the cumulant generating function by Eq.~(\ref{eigenvalue_eq}) has the symmetry $\lambda\to A_{\rm D}-\lambda$.
Therefore, the effective affinity reduces to the standard one
\be
\tilde A=A_{\rm D}=\beta (\mu_1-\mu_2)
\ee
if the QPC is at equilibrium.  This result can also be obtained from the definition (\ref{Aff_eff}) of the effective affinity.

\subsubsection{The limit $\vert T\vert\ll\vert\epsilon_{\rm A}-\epsilon_{\rm B}\vert$}

In this limit, the tunneling amplitude between the two dots composing the DQD is smaller than the difference between the energy levels of the dots, $\vert T\vert\ll\vert\epsilon_{\rm A}-\epsilon_{\rm B}\vert$.  Therefore, each dot is more strongly coupled to the next-neighboring reservoir than to the other dot.  Therefore, each dot equilibrates with the nearby reservoir on a time scale faster than for electron transfers.  There exist two subcases whether $\epsilon_{\rm A}-\epsilon_{\rm B}$ is positive or negative.

If $\epsilon_{\rm A}>\epsilon_{\rm B}$, the mixing angle goes to zero $\theta\to 0$ so that $\vert + \rangle \simeq \vert 1_{\rm A}0_{\rm B}\rangle$ and $\vert - \rangle \simeq -\vert 0_{\rm A}1_{\rm B}\rangle$.  In this subcase, the transition rates separate in the two groups:
\be
a_{1+},b_{1+},a_{2-},b_{2-} \gg a_{1-},b_{1-},a_{2+},b_{2+} ,c_{+-},c_{-+} = O(\theta^2)
\ee
Accordingly, the matrix (\ref{L_lambda}) splits in two as $\mbox{\helvb L}(\lambda)=\mbox{\helvb L}_0(\lambda)+\mbox{\helvb L}_1(\lambda)$ where $\mbox{\helvb L}_0=O(\theta^0)$ and $\mbox{\helvb L}_1=O(\theta^2)$ as $\theta\to 0$.  In Eq.~(\ref{eigenvalue_eq}), the eigenvector and the eigenvalue can be expanded similarly as ${\bf v}={\bf v}_0 + {\bf v}_1+\cdots$ and $Q=Q_0+Q_1+\cdots$.  At zeroth order, the eigenvalue vanishes $Q_0=0$, the right-hand eigenvector is given by
\be
{\bf v}_0=
\left(
\begin{array}{c}
1 \\
{\rm e}^{-\beta(\epsilon_+-\mu_1)}{\rm e}^{-\lambda} \\
{\rm e}^{-\beta(\epsilon_--\mu_2)} 
\end{array}
\right)
\ee
and the left-hand eigenvector such that ${\bf u}_0^{\rm T}\cdot\mbox{\helvb L}_0=0$ by
\be
{\bf u}_0^{\rm T} = \left( \ 1 \quad {\rm e}^{\lambda} \quad 1 \ \right)
\ee
At first order in $\theta^2$, we find that
\be
Q_1 \simeq - \frac{{\bf u}_0^{\rm T}\cdot\mbox{\helvb L}_1\cdot{\bf v}_0}{{\bf u}_0^{\rm T}\cdot{\bf v}_0}
\ee
which gives the leading approximation to the cumulant generating function $Q\simeq Q_1$:
\be
Q(\lambda)\simeq J_+\left(1-{\rm e}^{-\lambda}\right) + J_-\left(1-{\rm e}^{\lambda}\right)  
\label{Q_theta=0}
\ee
with
\bea
&&J_+= \frac{a_{1-}+(b_{2+}+c_{-+}){\rm e}^{-\beta(\epsilon_+-\mu_1)}}{1+{\rm e}^{-\beta(\epsilon_+-\mu_1)}+{\rm e}^{-\beta(\epsilon_--\mu_2)}}\\
&& J_-= \frac{a_{2+}+(b_{1-}+c_{+-}){\rm e}^{-\beta(\epsilon_--\mu_2)}}{1+{\rm e}^{-\beta(\epsilon_+-\mu_1)}+{\rm e}^{-\beta(\epsilon_--\mu_2)}}
\eea
The cumulant generating function (\ref{Q_theta=0}) has the symmetry (\ref{FT-Q}) of the fluctuation theorem with the effective affinity:
\be
\tilde A = \beta(\mu_1-\mu_2) + \ln\frac{a_{1-}{\rm e}^{-\beta\mu_1}+a_{2+}{\rm e}^{-\beta\mu_2}+c_{-+}{\rm e}^{-\beta\epsilon_+}}{a_{1-}{\rm e}^{-\beta\mu_1}+a_{2+}{\rm e}^{-\beta\mu_2}+c_{+-}{\rm e}^{-\beta\epsilon_-}}
\ee
We notice that, in the logarithm, the numerator and the denominator are both vanishing proportionally to $\theta^2$ so that their ratio is non vanishing in the limit $\theta\to 0$ and thus gives a contribution to the effective affinity beyond its standard value (\ref{A_D}).  This expression is equivalently obtained from Eq.~(\ref{Aff_eff}) in the limit $\theta\to 0$.

If $\epsilon_{\rm A}<\epsilon_{\rm B}$, the mixing angle has the limit $\theta\to \pi$ so that $\vert -\rangle \simeq \vert 1_{\rm A}0_{\rm B}\rangle$ and $\vert + \rangle \simeq \vert 0_{\rm A}1_{\rm B}\rangle$.
In this other subcase, the transition rates separate as
\be
a_{1-},b_{1-},a_{2+},b_{2+} \gg  a_{1+},b_{1+},a_{2-},b_{2-},c_{+-},c_{-+} = O(\delta^2)
\ee
with $\delta=\pi-\theta$.  Hence, the cumulant generating function can be obtained with a similar method as in the previous limit to get
\be
Q(\lambda)\simeq J'_+\left(1-{\rm e}^{-\lambda}\right) + J'_-\left(1-{\rm e}^{\lambda}\right)  
\label{Q_theta=pi}
\ee
with
\bea
&&J'_+= \frac{a_{1+}+(b_{2-}+c_{+-}){\rm e}^{-\beta(\epsilon_--\mu_1)}}{1+{\rm e}^{-\beta(\epsilon_--\mu_1)}+{\rm e}^{-\beta(\epsilon_+-\mu_2)}}\\
&& J'_-= \frac{a_{2-}+(b_{1+}+c_{-+}){\rm e}^{-\beta(\epsilon_+-\mu_2)}}{1+{\rm e}^{-\beta(\epsilon_--\mu_1)}+{\rm e}^{-\beta(\epsilon_+-\mu_2)}}
\eea
Here also, the symmetry (\ref{FT-Q}) of the fluctuation theorem is satisfied by the generating function (\ref{Q_theta=pi}) but with the effective affinity:
\be
\tilde A = \beta(\mu_1-\mu_2) + \ln\frac{a_{1+}{\rm e}^{-\beta\mu_1}+a_{2-}{\rm e}^{-\beta\mu_2}+c_{+-}{\rm e}^{-\beta\epsilon_-}}{a_{1+}{\rm e}^{-\beta\mu_1}+a_{2-}{\rm e}^{-\beta\mu_2}+c_{-+}{\rm e}^{-\beta\epsilon_+}}
\label{Aff_eff_theta=pi}
\ee
which is equivalently given by Eq.~(\ref{Aff_eff}) in the limit $\theta\to \pi$.
In the logarithm, the numerator and the denominator are both vanishing proportionally to $\delta^2=(\pi-\theta)^2$ so that their ratio is non vanishing in the limit $\theta\to \pi$ and here also modifies the effective affinity with respect to its standard value (\ref{A_D}).

\begin{figure}[htbp]
\centerline{\includegraphics[width=9cm]{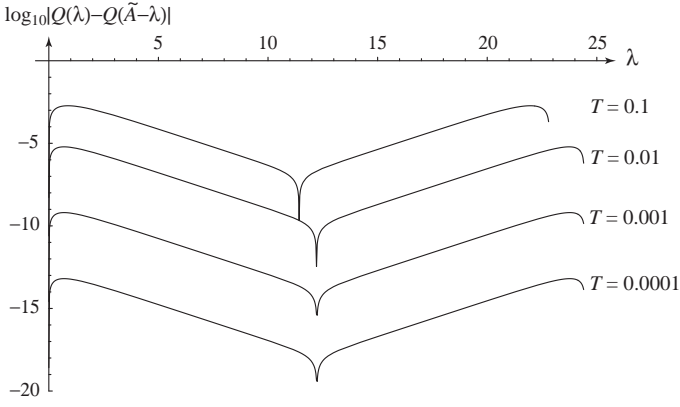}}
\caption{The difference between both sides of Eq.~(\ref{FT-Q}) in absolute value versus the counting parameter $\lambda$ for smaller and smaller values of the tunneling amplitude: $T=0.1$, $T=0.01$, $T=0.001$, and $T=0.0001$.  The other parameters are $\mu_1=3$, $\mu_2=1$, $\Delta\mu_{\rm C}=5$, $\beta =10$, $\epsilon_{\rm A} = 0.7$, $\epsilon_{\rm B}  = 1.2$, $T_{\rm A} = T_{\rm B} = 0.1$, $U_{\rm A} =-0.21$, $U_{\rm B} = -0.06$, $\kappa = 0.2$, and $g = 1$.  For $T=0.0001$, the effective affinity is equal to $\tilde A=24.4594$ and the mixing angle to $\theta=\pi-0.0004$.  In the limit $T=0$, the effective affinity (\ref{Aff_eff_theta=pi}) is equal to $\tilde A=\beta(\mu_1-\mu_2)+\Delta\tilde A$ with $A_{\rm D}=\beta(\mu_1-\mu_2)=20$ and $\Delta\tilde A=4.45945$.}
\label{fig8}
\end{figure}

Figure~\ref{fig8} shows that the difference between both sides of Eq.~(\ref{FT-Q}) is indeed vanishing as the tunneling amplitude $T$ gets smaller and smaller so that  the symmetry (\ref{FT-Q}) is indeed satisfied in the limit $\theta\to\pi$.

\subsubsection{The QPC induces fast transitions $\vert + \rangle \leftrightharpoons \vert - \rangle$}

Here, we consider the limit
\be
\vert U_{\rm A}-U_{\rm B}\vert \gg \vert T_{\rm A}\vert, \vert T_{\rm B}\vert
\ee
so that the rates of the transitions $\vert + \rangle \leftrightharpoons \vert - \rangle$ are larger than the other rates:
\be
c_{+-}, c_{-+} \gg a_{js}, b_{js}
\label{limit_red}
\ee
with $j=1,2$ and $s=\pm$.  In this case, there is no distinction between the states $\vert\pm\rangle$ on the intermediate time scale $\Delta t$ between the short time of the transitions $\vert + \rangle \leftrightharpoons \vert - \rangle$  and the dwell time of electrons in the DQD: $c_{\pm,\mp}^{-1}\ll\Delta t\ll a_{js}^{-1}, b_{js}^{-1}$.  Therefore, the stochastic process admits a reduced description in terms of the probability 
\be
p_1(n,t) = p_+(n,t)+p_-(n,t)
\ee
that the DQD is occupied and the probability $p_0(n,t)$ that it is empty.  The probabilities of the states $\vert\pm\rangle$ are obtained as
\be
p_{\pm}(n,t) = p_1(n,t) \, P_{\pm\vert 1}
\ee
in terms of the conditional probabilities of the states $\vert\pm\rangle$ given that the DQD is occupied:
\bea
&& P_{+\vert 1} = \frac{c_{+-}}{c_{+-}+c_{-+}} \\
&& P_{-\vert 1} = \frac{c_{-+}}{c_{+-}+c_{-+}} 
\eea
These conditional probabilities are normalized according to $P_{+\vert 1}+P_{-\vert 1}=1$.
The master equation (\ref{master}) thus reduces to
\be
\left(
\begin{array}{c}
\partial_t\, p_{0}(n,t) \\
\partial_t\, p_{1}(n,t) 
\end{array}
\right)
=
\left(
\begin{array}{cc}
-a_{1}-a_{2} & b_{1}\,\hat{E}^+ +b_{2} \\
a_{1}\,\hat{E}^- +a_{2} & -b_{1}-b_{2}
\end{array}
\right)
\left(
\begin{array}{c}
p_{0}(n,t) \\
p_{1}(n,t) 
\end{array}
\right)
\label{L-01}
\ee
with the coefficients
\bea
&& a_j \equiv a_{j+}+a_{j-} \\
&& b_j \equiv b_{j+} P_{+\vert 1}+b_{j-} P_{-\vert 1} = \frac{b_{j+}c_{+-}+b_{j-}c_{-+}}{c_{+-}+c_{-+}}
\eea
for $j=1,2$.

The cumulant generating function is here given by
\begin{widetext}
\be
Q(\lambda) =  \frac{1}{2} \left[ 
a_1+a_2 +b_1+b_2 -
\sqrt{\left(a_1+a_2 -b_1-b_2 \right)^2
+ 4 \left( a_1\,{\rm e}^{-\lambda}+a_2\right)\left(b_1\,{\rm e}^{+\lambda}+b_2\right)}\right]
\label{Q_red}
\ee
\end{widetext}
The symmetry (\ref{FT-Q}) is again satisfied with the effective affinity:
\be
\tilde A= \ln \frac{a_1b_2}{a_2 b_1} = \ln\frac{(a_{1+}+a_{1-})\left(b_{2+}c_{+-}+b_{2-}c_{-+}\right)}{(a_{2+}+a_{2-})\left(b_{1+}c_{+-}+b_{1-}c_{-+}\right)}
\label{Aff_eff_red}
\ee
which can be obtained from Eq.~(\ref{Aff_eff}) in the limit (\ref{limit_red}).

\begin{figure}[h]
\centerline{\includegraphics[width=9cm]{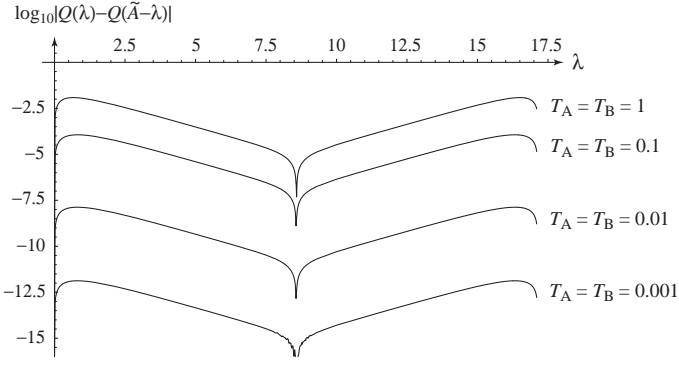}}
\caption{The difference between both sides of Eq.~(\ref{FT-Q}) in absolute value versus the counting parameter $\lambda$ for smaller and smaller values of the tunneling amplitudes $T_{\rm A} = T_{\rm B}$.  The other parameters are $\mu_1=3$, $\mu_2=1$, $\Delta\mu_{\rm C}=5$, $\beta =10$, $\epsilon_{\rm A} = 0.7$, $\epsilon_{\rm B}  = 1.2$, $T=0.5$, $U_{\rm A} =-0.21$, $U_{\rm B} = -0.06$, $\kappa = 0.2$, and $g = 1$.  For $T_{\rm A} = T_{\rm B}=0.001$, the effective affinity is equal to $\tilde A=17.1397$.}
\label{fig9}
\end{figure}

Figure~\ref{fig9} shows that the symmetry (\ref{FT-Q}) is well satisfied in the limit (\ref{limit_red}) as the tunneling amplitudes $T_{\rm A}$ and $T_{\rm B}$ between the DQD and the reservoirs $j=1,2$ are decreased.

\subsection{Dependence of the effective affinity on the quantum dot energies}

The gate voltages applied to the quantum dots control their energy levels.  Therefore, varying the energies $\epsilon_{\rm A}$ and $\epsilon_{\rm B}$ corresponds in the present model to changing the gate voltages of the quantum dots A and B.

\begin{figure}[h]
\centerline{\includegraphics[width=9cm]{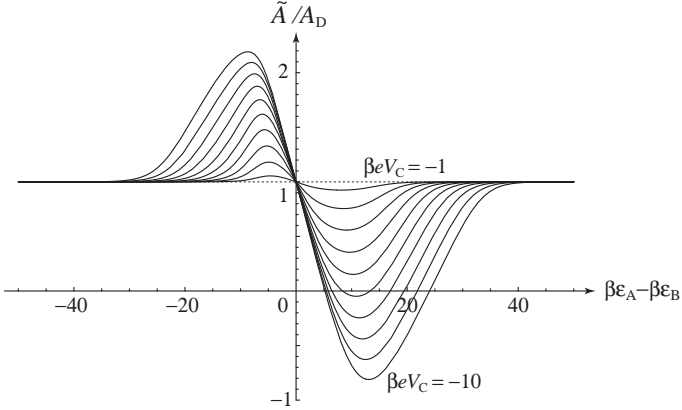}}
\caption{The effective affinity (\ref{Aff_eff}) rescaled by the actual affinity (\ref{A_D}) versus the energy difference $\beta\epsilon_{\rm A}-\beta\epsilon_{\rm B}$ in the DQD for several values of the potential difference applied to the QPC: $\beta\Delta\mu_{\rm C}=\beta eV_{\rm C}=-1,-2,-3,...,-10$.  The other parameters are $\beta\epsilon_{\rm A}+\beta\epsilon_{\rm B}=5$, $\beta\mu_1=2$, $\beta\mu_2=-2$, $T=0.1$, $T_{\rm A} = T_{\rm B}=0.1$, $U_{\rm A} =-1.2$, $U_{\rm B} =-1.8$, $\kappa = 0.2$, and $g = 1$.  The effective affinity is equal to the value $A_{\rm D}$ at $\beta\epsilon_{\rm A}-\beta\epsilon_{\rm B}\simeq-0.00367$ in the present conditions.}
\label{fig10}
\end{figure}

In Fig.~\ref{fig10}, the effective affinity (\ref{Aff_eff}) is depicted as a function of the energy difference $\epsilon_{\rm A}-\epsilon_{\rm B}$ in the DQD for several values of the affinity (\ref{A_C}) in the QPC circuit.  The backaction of the QPC onto the DQD circuit manifests itself by the deviations of the ratio $\tilde A/A_{\rm D}$ from unity.  As expected, the backaction gets larger as the QPC is driven further away from equilibrium by increasing the absolute value of its affinity.  Although the effective affinity of the DQD is nearly equal to its actual value (\ref{A_D}) if the QPC is close to equilibrium for $\Delta\mu_{\rm C}=-1$, they may significantly differ from each other if the QPC is far from equilibrium.  Under some conditions, the sign of the effective affinity may even be reversed with respect to the actual value (\ref{A_D}).  We notice that the average current in the DQD should also change its sign under these conditions, because the inequality (\ref{inequal}) is always satisfied.  This change of sign of the DQD current is the consequence of the Coulomb drag effect due to the QPC.
In this regime, the thermodynamic efficiency (\ref{eta}) is bounded according to Eq.~(\ref{eta-bound}).  

In Fig.~\ref{fig10}, we also observe that the effective affinity $\tilde A$ converges to its basic value $A_{\rm D}$ for $\vert\epsilon_{\rm A}-\epsilon_{\rm B}\vert \gg \vert\Delta \mu_{\rm C}\vert$.   The reason is that, in this limit, the rates of the transitions populating the state $\vert + \rangle$ are vanishing: $a_{1+}=a_{2+}=c_{+-}=0$.  Accordingly, the state $\vert + \rangle$ is never populated and it gets out of the dynamics ruled by the master equation (\ref{master-m})-(\ref{L}): $\lim_{t\to\infty}p_+(n,t)=0$.  In this case, the effective affinity (\ref{Aff_eff}) becomes $\tilde A = \ln(a_{1-}b_{2-})/(b_{1-}a_{2-})=\beta(\mu_1-\mu_2)=A_{\rm D}$.  We point out that, if each quantum dot had more than the sole energy level assumed in the present model, the effective affinity would become more complicated for $\vert\epsilon_{\rm A}-\epsilon_{\rm B}\vert \gg \vert\Delta \mu_{\rm C}\vert$.

\begin{figure}[h]
\centerline{\includegraphics[width=8cm]{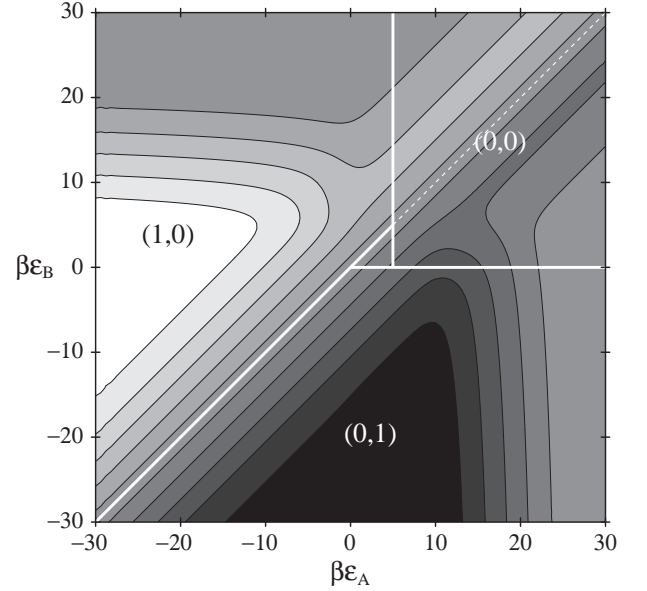}}
\caption{The effective affinity (\ref{Aff_eff}) rescaled by the actual affinity (\ref{A_D}) versus the energies $\beta\epsilon_{\rm A}$ and $\beta\epsilon_{\rm B}$ of the quantum dots.  The parameter values are $\beta\mu_1=5$, $\beta\mu_2=0$, $\beta\Delta\mu_{\rm C}=14$, $T=0.1$, $T_{\rm A} = T_{\rm B}=0.1$, $U_{\rm A} =-0.06$, $U_{\rm B} =-0.15$, $\kappa = 0.05$, and $g = 1$.  The white lines are the borders between the domains where the quantum dots are dominantly occupied according to $(n_{\rm A},n_{\rm B})\simeq(0,0),(0,1),(1,0)$.  We notice that the effective affinity is equal to the actual one, $\tilde A=A_{\rm D}$, along the diagonal $\beta\epsilon_{\rm A}=\beta\epsilon_{\rm B}$.  The effective affinity reaches the value $\tilde A/A_{\rm D}\simeq 3.5$ in the white area and $\tilde A/A_{\rm D}\simeq -1.5$ in the dark area.  The values of the parameters $U_{\rm A}/U_{\rm B}$ and $\kappa$ are estimated from experimental conditions.\cite{FHTH06}  In the present conditions, the effective affinity $\tilde A$ is equal to $A_{\rm D}$ near $\beta\epsilon_{\rm B}\simeq \beta\epsilon_{\rm A}+0.00024$ at $\beta\epsilon_{\rm A}=-10$, near $\beta\epsilon_{\rm B}\simeq \beta\epsilon_{\rm A}+0.00099$ at $\beta\epsilon_{\rm A}=0$, and near $\beta\epsilon_{\rm B}\simeq \beta\epsilon_{\rm A}+0.00166$ at $\beta\epsilon_{\rm A}=10$.}
\label{fig11}
\end{figure}

Figure~\ref{fig11} shows the ratio of the effective affinity (\ref{Aff_eff}) to the actual affinity (\ref{A_D}) in the plane of the quantum dot energies $(\epsilon_{\rm A},\epsilon_{\rm B})$ in comparison with the domains where the quantum dots have the dominant occupancies $(0,0)$, $(0,1)$, or $(1,0)$.\cite{WFEFTK03}  These domains are delimited by three straight lines shown in Fig.~\ref{fig11}.  The DQD is empty in the domain where $\epsilon_{\rm A}>\mu_1$ and $\epsilon_{\rm B}>\mu_2$.  The dot A is empty while the dot B is occupied by one electron in the domain where $\epsilon_{\rm A}>\epsilon_{\rm B}$ and $\epsilon_{\rm B}<\mu_2$. The dot A is occupied by one electron while the dot B is empty in the domain where $\epsilon_{\rm A}<\mu_1$ and $\epsilon_{\rm B}>\epsilon_{\rm A}$.  In the triangle between these three domains, the DQD is excited in its upper state $\vert +\rangle$ and its current is maximal, as expected.\cite{WFEFTK03}  The effective affinity takes its actual value $\tilde A=A_{\rm D}$ very close to the diagonal $\epsilon_{\rm A}=\epsilon_{\rm B}$, which explains that the frontier between the domains is a favorable region to minimize the deviation of the effective affinity with respect to the value (\ref{A_D}) fixed by the voltage across the DQD.\cite{KRBMGUIE12}  Besides, the effective affinity (\ref{Aff_eff}) can significantly differs from its actual value (\ref{A_D}).  In Fig.~\ref{fig11}, the effective affinity ranges from $\tilde A\simeq 3.5\times A_{\rm D}$ in the white area down to $\tilde A\simeq -1.5\times A_{\rm D}$ in the dark area.  In particular, the effective affinity drops from its actual value in the domain $(0,1)$ away from the frontier between $(0,1)$ and $(1,0)$.

\section{Summary and conclusions}
\label{Summary}

The present paper reports the study of electronic transport properties in a DQD circuit capacitively coupled to a QPC.  The QPC plays the role of detector for the single-electron transfers in the DQD and also affects this latter because of the backaction due to its noise.

The system is modeled by a simple Hamiltonian capturing the main features of such circuits.  The reservoirs in contact with the DQD as well as the QPC itself are described by tight-binding quadratic Hamiltonians that are exactly solvable, allowing the non-perturbative analysis of the QPC in arbitrary nonequilibrium states.  The tunneling between the DQD and its reservoirs, as well as the capacitive coupling between the DQD and the QPC are treated at second order of perturbation theory together with the rotating-wave and the wide-band approximations.  The electron transport in the QPC is supposed to behave faster than in the DQD.  The double occupancy of the DQD is assumed to lie at high enough energy to be neglected.

In this way, a Markovian master equation is obtained for the stochastic process of electron transfers across the DQD.  This master equation holds for the QPC in regimes arbitrarily far from equilibrium.
Under the assumption that the DQD is slower than the QPC, the current in the QPC is described by a Landauer-B\"uttiker formula depending on the time-dependent quantum state of the DQD.

The asymmetry of the capacitive coupling required for the bidirectional counting of electron transfers with the QPC is also responsible for its backaction onto the DQD current.  This backaction induces transitions between the internal states of the occupied DQD and, consequently, the Coulomb drag of the DQD current by the QPC if this latter is out of equilibrium.  Remarkably, a current is induced in the DQD if the voltage applied to the QPC exceeds a threshold given by the internal energies of the DQD, which is consistent with experimental observations.\cite{GSLIEDG07}

On the basis of the master equation, the FCS is established for electron transport in the DQD.  Thanks to  its cumulant generating function, an effective affinity is introduced that characterizes the nonequilibrium driving of the DQD not only by the voltage applied to it, but also by the capacitively coupled QPC.
The value of this effective affinity differs from the value fixed by the voltage bias across the DQD if the QPC is driven out of equilibrium.

In the present paper, our main result is the establishment of effective fluctuation theorems for the DQD current under specific conditions.  On fundamental ground, a bivariate fluctuation theorem is known to hold for both DQD and QPC currents.  However, the sole current across the DQD does not generally obey a fluctuation theorem because of the capacitive coupling to the QPC.  Therefore, it is surprising that there exist conditions under which effective fluctuation theorems can nevertheless be established for the sole DQD current.  Our analysis shows that a single-current fluctuation theorem is valid under every one of the following conditions:

(1) If the QPC is at equilibrium, in which case the effective affinity remains equal to its value fixed by the voltage applied to the DQD.

(2) If the tunneling amplitude $T$ between the quantum dots composing the DQD is smaller than the difference between their internal energies $\epsilon_{\rm A}$ and $\epsilon_{\rm B}$: $\vert T\vert \ll \vert \epsilon_{\rm A}-\epsilon_{\rm B}\vert$.  In this limit, each dot is essentially at equilibrium with its next-neighboring reservoir and the DQD circuit behaves as another quantum point contact.

(3) If the asymmetry of the capacitive coupling to the QPC is stronger than the tunneling of the DQD to its reservoirs: $\vert U_{\rm A}-U_{\rm B}\vert \gg \vert T_{\rm A}\vert, \vert T_{\rm B}\vert$.  In this case, the QPC induces transitions between the two single-electron internal states of the DQD that are faster than its charging or discharging.  Therefore, the stochastic description reduces to a process for only two internal states: the empty and the singly occupied states.  

Moreover, the effective affinity is analyzed for its dependence on the internal energies of the quantum dots.  In the present model, they constitute the control parameters that are analogue to the gate voltages of the quantum dots.  Interestingly, the effective affinity is shown to remain close to the DQD voltage bias at the frontier between the domains of single occupancy of the DQD $(n_{\rm A},n_{\rm B})\simeq(0,1),(1,0)$ and to deviate from this value away from this frontier.

Besides, in the regimes where an effective fluctuation theorem holds and the Coulomb drag may reverse the DQD current, the thermodynamic entropy production turns out to have a positive lower bound equal to the product of the effective affinity with the average DQD current.  In these regimes, the thermodynamic efficiency of electron pumping by Coulomb drag is limited by an upper bound lower than unity in terms of the ratio of the effective affinity to the voltage applied across the DQD.

To conclude, the effective affinity can be directly measured if a single-current fluctuation theorem is observed to hold experimentally.  Under such circumstances, the effective affinity can be used to characterize the FCS of electron transport and the mechanisms driving the circuit out of thermodynamic equilibrium.

\begin{acknowledgments}
G. Bulnes Cuetara thanks the ``Fonds pour la Formation \`a la Recherche dans l'Industrie et l'Agriculture" (FRIA Belgium) for financial support.  M. Esposito is supported by the National Research Fund, Luxembourg in the frame of the project FNR/A11/02.  G. Schaller acknowledges support by the DFG (SCHA~1646/2-1).  This research is also supported by the Belgian Federal Government under the Interuniversity Attraction Pole project P7/18 ``DYGEST".
\end{acknowledgments}

\appendix

\section{Diagonalization of the DQD Hamiltonian}
\label{AppA}

The diagonalization of the DQD Hamiltonian (\ref{H_DQD}) can be performed analytically.  The states of the local basis with one charge are defined as
\bea
&&\vert 1_{\rm A}0_{\rm B}\rangle \equiv d_{\rm A}^{\dagger} \vert 0_{\rm A}0_{\rm B}\rangle \\
&&\vert 0_{\rm A}1_{\rm B}\rangle \equiv d_{\rm B}^{\dagger} \vert 0_{\rm A}0_{\rm B}\rangle
\eea
where $\vert 0_{\rm A}0_{\rm B}\rangle$ is the ground state of the DQD.  Discarding the double-occupancy state, the eigenstates are thus expressed as
\bea
&&\vert 0 \rangle = \vert 0_{\rm A}0_{\rm B}\rangle\\
&&\vert + \rangle = \cos\frac{\theta}{2} \; \vert 1_{\rm A}0_{\rm B}\rangle + \sin\frac{\theta}{2} \; \vert 0_{\rm A}1_{\rm B}\rangle \\
&&\vert - \rangle = \sin\frac{\theta}{2} \; \vert 1_{\rm A}0_{\rm B}\rangle - \cos\frac{\theta}{2} \; \vert 0_{\rm A}1_{\rm B}\rangle 
\eea
with the mixing angle
\be
\tan\theta = \frac{2T}{\epsilon_{\rm A}-\epsilon_{\rm B}}
\ee
The corresponding eigenvalues are given by
\bea
&&\epsilon_0 = 0 \\
&&\epsilon_{\pm} = \frac{\epsilon_{\rm A}+\epsilon_{\rm B}}{2} \pm \sqrt{\left(\frac{\epsilon_{\rm A}-\epsilon_{\rm B}}{2}\right)^2 + T^2} 
\eea
If $T$ vanishes, the mixing angle goes to $\theta=0$ if $\epsilon_{\rm A}>\epsilon_{\rm B}$ and to $\theta=\pi$ if $\epsilon_{\rm B}>\epsilon_{\rm A}$.

In the eigenbasis, the tunneling amplitudes between the DQD and its reservoirs are obtained as
\bea
&&T_{1+} = T_{\rm A} \, \cos\frac{\theta}{2} \label{T1+}\\
&&T_{1-} = T_{\rm A} \, \sin\frac{\theta}{2} \label{T1-}\\
&&T_{2+} = T_{\rm B} \, \sin\frac{\theta}{2} \label{T2+}\\
&&T_{2-} = -T_{\rm B} \, \cos\frac{\theta}{2} \label{T2-}
\eea
and the capacitive coupling parameters with the QPC as
\bea
&&U_{++} = \frac{1}{2}(U_{\rm A}+U_{\rm B}) + \frac{1}{2}(U_{\rm A}-U_{\rm B}) \cos\theta \label{U++}\\
&&U_{+-} = U_{-+}=  \frac{1}{2}(U_{\rm A}-U_{\rm B}) \sin\theta \label{U+-} \\
&&U_{--} = \frac{1}{2}(U_{\rm A}+U_{\rm B}) - \frac{1}{2}(U_{\rm A}-U_{\rm B}) \cos\theta \label{U--}
\eea
We notice that $U_{+-} = U_{-+}=0$ at $\theta=0$ and $\theta=\pi$.

\section{Diagonalization of the reservoir Hamiltonians}
\label{AppB}

The tight-binding Hamiltonian operators (\ref{H_j}) of the reservoirs can be diagonalized into\cite{T01,S07}
\be
H_j = \int_0^{\pi} dk \, \epsilon_k \, c_{j,k}^{\dagger} c_{j,k} \qquad (j=1,2,3,4)
\label{H_j_diag}
\ee
with the energy eigenvalues
\be
\epsilon_k = - 2 \gamma \, \cos k \qquad (0\leq k \leq \pi)
\label{E_k}
\ee
The new annihilation operators are related to the previous ones by
\bea
&& c_{j,k}= \sum_{l=0}^{\infty}  \phi_k(l) \, d_{j,l}\\
&& d_{j,l}=\int_0^{\pi} dk \, \phi_k(l) \, c_{j,k} \label{dc}
\eea
in terms of the real eigenfunctions
\be
\phi_k(l) = \sqrt{\frac{2}{\pi}} \, \sin k (l+1) \qquad (l=0,1,2,3,...)
\label{phi_k}
\ee
forming a complete orthonormal basis
\bea
&& \int_0^{\pi} dk\, \phi_k(l)\, \phi_k(l') = \delta_{ll'} \\
&& \sum_{l=0}^{\infty}  \phi_k(l) \, \phi_{k'}(l) = \delta(k-k')
\eea

The creation-annihilation operators anticommute to express the fermionic character of the electrons.
The annihilation operators have the following free time evolution:
\be
c_{j,k}(t) = {\rm e}^{iH_jt} c_{j,k} {\rm e}^{-iH_jt} =c_{j,k} \, {\rm e}^{-i\epsilon_kt} 
\label{c_t}
\ee
Similarly, the number operator (\ref{N_j}) is diagonalized into
\be
N_j = \int_0^{\pi} dk \, c_{j,k}^{\dagger} c_{j,k}
\label{N_j_diag}
\ee

If a reservoir is composed of $L$ sites of indices $0\leq l \leq L-1$, the wavenumber takes the discrete values $k=n\pi/L$ with $n=1,2,3,...,L$ separated by $\Delta k =\pi/L$ and the density of states is given by
\be
D(\epsilon)=\sum_k \delta(\epsilon-\epsilon_k) =\frac{L}{\pi\sqrt{4\gamma^2-\epsilon^2}} 
\ee
The average local density of states is thus equal to
\be
g(\epsilon)=\frac{1}{L}\, D(\epsilon) =\frac{1}{\pi\sqrt{4\gamma^2-\epsilon^2}} 
\ee
The energy band extends over the interval $-2\gamma \leq \epsilon\leq +2\gamma$ and the average local density of states in the middle of the band is given by
\be
g\equiv g(0) = \frac{1}{2\pi\gamma} 
\label{LDOS}
\ee

A large reservoir in the grand-canonical equilibrium ensemble at the inverse temperature $\beta$ and the chemical potential $\mu_j$ is described by the density operator
\be
\rho_j = \frac{1}{\Xi_j} \, {\rm e}^{-\beta(H_j-\mu_jN_j)}
\label{rho_j}
\ee
where the partition function $\Xi_j$ guarantees the normalization condition ${\rm tr}\, \rho_j=1$.
In this statistical ensemble, the quadratic combinations of the creation-annihilation operators have the statistical averages
\bea
&&\langle c_{j,k}^{\dagger} \, c_{j,k'}\rangle = f_{jk} \; \delta(k-k') \label{<c+c>}\\
&&\langle c_{j,k} \, c_{j,k'}^{\dagger}\rangle = \left(1-f_{jk}\right) \, \delta(k-k') \label{<cc+>}
\eea
where $f_{jk} = f_j(\epsilon_k)$ is the Fermi-Dirac distribution (\ref{FD}).

\section{Diagonalization of the QPC Hamiltonian}
\label{AppC}

It is supposed that there is no bound state in the QPC, which requires that $\vert T_{\rm C}\vert < \gamma$.  The diagonalization of the QPC Hamiltonian (\ref{H_C}) is solved as a scattering problem in which the point contact between the reservoirs $j=3$ and $j=4$ is the scatterer.\cite{T01,S07}  Since the Hamiltonian and the particle number are quadratic, they can be transformed into
\bea
&& H_{\rm C} = \int_{-\pi}^{+\pi} dq \, \epsilon_q \, c_{q}^{\dagger} c_{q}
\label{H_C_diag} \\
&& N_{\rm C} = \int_{-\pi}^{+\pi} dq \, c_{q}^{\dagger} c_{q}
\label{N_C_diag} 
\eea
with the energy eigenvalues
\be
\epsilon_q = - 2 \gamma \, \cos q \qquad (-\pi\leq q \leq +\pi)
\label{E_q}
\ee
with $\gamma >0$.
The annihilation operators are transformed according to
\bea
&& c_{q}= \sum_{j=3,4} \sum_{l=0}^{\infty}  \psi_q^*(j,l) \, d_{j,l}\\
&& d_{j,l}=\int_{-\pi}^{+\pi} dq \, \psi_q(j,l) \, c_{q}
\eea
in terms of the scattering eigenfunctions
\bea
\left\{
\begin{array}{l}
\psi_q(3,l) = \frac{1}{\sqrt{2\pi}}\left({\rm e}^{-iql}+r_q\, {\rm e}^{iql}\right) \\
\psi_q(4,l) = \frac{1}{\sqrt{2\pi}}\; t_q\, {\rm e}^{iql}
\end{array}
\right. \label{psi+}\\
\left\{
\begin{array}{l}
\psi_{-q}(3,l) = \frac{1}{\sqrt{2\pi}}\; t_q\, {\rm e}^{iql}\\
\psi_{-q}(4,l) = \frac{1}{\sqrt{2\pi}}\left({\rm e}^{-iql}+r_q\, {\rm e}^{iql}\right) 
\end{array}
\right. \label{psi-}
\eea
for $q>0$ and $l=0,1,2,3,...$.  The transmission amplitude is given by
\be
t_q = - T_{\rm C} \, \gamma\, \frac{{\rm e}^{iq}-{\rm e}^{-iq}}{T_{\rm C}^2-\gamma^2{\rm e}^{-2iq}}
\ee
and it is related to the reflection amplitude by
\be
1+r_q= -\frac{\gamma}{T_{\rm C}} \, t_q \, {\rm e}^{-iq}
\ee
which determines the transmission probability
\be
{\cal T}_{\epsilon} = \vert t_{q(\epsilon)}\vert^2 = \frac{T_{\rm C}^2\, (4\gamma^2-\epsilon^2)}{(T_{\rm C}^2+\gamma^2)^2-T_{\rm C}^2\epsilon^2}
\label{T_E}
\ee
The transmission probability is maximal in the middle of the band where it reaches the value
\be
{\cal T}_{0} = \vert t_0\vert^2= \frac{4\, T_{\rm C}^2\, \gamma^2}{(T_{\rm C}^2+\gamma^2)^2}
\label{T_0}
\ee
and it vanishes at the edges of the energy band.  Since the band width $\Delta\epsilon=4\gamma$ is related to the local density of states in the middle of the band by Eq.~(\ref{LDOS}), the transmission probability can be written as
\be
{\cal T}_0 = \frac{4\kappa}{(1+\kappa)^2}
\ee
in terms of the dimensionless contact transparency \cite{GK06}
\be
\kappa = (2\pi \, g \, T_{\rm C})^2 = (T_{\rm C}/\gamma)^2
\label{kappa}
\ee
We notice that the transparency satisfies $\kappa < 1$ because of the condition $\vert T_{\rm C}\vert <\gamma$, which is required for the absence of bound state.

Under this condition, the scattering eigenfunctions (\ref{psi+})-(\ref{psi-}) form a complete orthonormal basis
\bea
&& \int_{-\pi}^{+\pi} dq\, \psi_q(j,l)\, \psi_q^*(j',l') = \delta_{jj'}\, \delta_{ll'} \\
&& \sum_{j=3,4}\sum_{l=0}^{\infty}  \psi_q(j,l) \, \psi_{q'}^*(j,l) = \delta(q-q')
\eea

Here, the annihilation operators have the free time evolution:
\be
c_{q}(t) = {\rm e}^{iH_{\rm C}t} c_{q} {\rm e}^{-iH_{\rm C}t} =c_{q} \, {\rm e}^{-i\epsilon_qt} 
\ee

The Hamiltonian operator (\ref{H_C_diag}) splits as
\be
H_{\rm C}= H_{\rm C}^{(+)} + H_{\rm C}^{(-)} 
\ee
 into the operators
\bea
&& H_{\rm C}^{(+)} = \int_{0}^{+\pi} dq \, \epsilon_q \, c_{q}^{\dagger} c_{q} \label{HC+}\\
&& H_{\rm C}^{(-)} = \int_{-\pi}^{0} dq \, \epsilon_q \, c_{q}^{\dagger} c_{q} \label{HC-}
\eea
These Hamiltonian operators commute
\be
[H_{\rm C}^{(+)},H_{\rm C}^{(-)}]=0
\ee
thanks to the diagonalization.  A similar decomposition holds for the particle number: $N_{\rm C}= N_{\rm C}^{(+)} + N_{\rm C}^{(-)}$.

If both reservoirs coupled by the QPC extended over $L$ sites of indices $0\leq l \leq L-1$, the wavenumber $q$ would take discrete values separated by $\Delta q =\pi/L$.
Accordingly, a nonequilibrium steady state for the QPC could be defined with the density operator
\be
\rho_{\rm C} = \frac{1}{\Xi_{\rm C}^{(+)}} \, {\rm e}^{-\beta(H_{\rm C}^{(+)}-\mu_3N_{\rm C}^{(+)})}
\; \frac{1}{\Xi_{\rm C}^{(-)}} \, {\rm e}^{-\beta(H_{\rm C}^{(-)}-\mu_4N_{\rm C}^{(-)})}
\label{rho_C}
\ee
properly normalized by the condition ${\rm tr}\, \rho_{\rm C}=1$.\cite{T01,LS11}
In this statistical ensemble, the quadratic combinations of the creation-annihilation operators have the statistical averages
\bea
&&\langle c_{q}^{\dagger} \, c_{q'}\rangle = f_{jq} \; \delta(q-q') \\
&&\langle c_{q} \, c_{q'}^{\dagger}\rangle = \left(1-f_{jq}\right) \, \delta(q-q')
\eea
with $j=3$ for $q>0$, $j=4$ for $q<0$, and the notation $f_{jq}=f_j(\epsilon_q)$ for the Fermi-Dirac distribution (\ref{FD}) at the inverse temperature $\beta$, the chemical potential $\mu_j$, and the wavenumber $q$.

\section{Derivation of the master equation}
\label{AppD}

\subsection{Generalities}

The interaction operator (\ref{V}) can be written as the sum
\be
V=\sum_{\alpha} S_{\alpha} \, R_{\alpha}
\label{V_sum}
\ee
where the operators $S_{\alpha}$ act on the subsystem degrees of freedom and the operators $R_{\alpha}$ on the environment of the subsystem.  The operators $S_{\alpha}$ and $R_{\alpha}$ commute or anticommute if they are linear or quadratic in the fermionic creation-annihilation operators.

Using perturbation theory at second order in the interaction (\ref{V_sum}), the rotating-wave approximation, and the Markovian limit, the master equation (\ref{master}) is obtained.
The rate of the transition $s'\to s$ involving the transfer of $\nu$ electrons is given by
\be
L_{ss'}^{(\nu)} = \sum_{\alpha\beta:\nu} \tilde C_{\alpha\beta}(\omega_{ss'}) \, \langle s \vert S_{\beta} \vert s'\rangle\langle s'\vert S_{\alpha}\vert s\rangle
\ee
where the sum extends over the terms in Eq.~(\ref{V_sum}) contributing to the transition.  

The transition rates are determined by the spectral functions
\be
\tilde C_{\alpha\beta}(\omega) = \int_{-\infty}^{+\infty} dt \;  C_{\alpha\beta}(t) \, {\rm e}^{-i\omega t}
\ee
defined as the Fourier transforms of the time-dependent correlation functions of the environment coupling operators:
\be
C_{\alpha\beta}(t) = \varsigma_{\alpha}\, \left\langle{\rm e}^{iH_{\rm R}t} \, \tilde R_{\alpha} \, {\rm e}^{-iH_{\rm R}t} \, \tilde R_{\beta}\right\rangle_{\rm R}
\label{C_ab}
\ee
with
\be
\tilde R_{\alpha}=R_{\alpha}-\langle R_{\alpha}\rangle_{\rm R}
\ee
and $\varsigma_{\alpha}=\pm 1$ whether the operators $S_{\alpha}$ and $R_{\alpha}$ commute or anticommute.  The statistical average in Eq.~(\ref{C_ab}) is taken over the initial density operator (\ref{rho_R0}) for the environment: 
\be
\langle\cdot\rangle_{\rm R}={\rm tr}_{\rm R} \, \rho_{\rm R}(0) \, (\cdot)
\ee

\subsection{Electron tunneling between the DQD and its reservoirs}

The interactions (\ref{V1A}) and (\ref{V2B}) describe electron tunneling between the DQD and the reservoirs $j=1,2$.  In the DQD eigenbasis, these interactions take the form given by Eqs.~(\ref{V1A_bis}) and (\ref{V2B_bis}) showing that they are linear in the creation-annihilation operators of the reservoirs.  Therefore, their average over the reservoir equilibrium ensemble is vanishing.

The charging rates (\ref{ajs}) appear in the following elements of the matrix (\ref{L}):
\be
a_{1+}=L_{+0}^{(-)} \; , \quad a_{1-}=L_{-0}^{(-)} \; , \quad a_{2+}=L_{+0}^{(0)} \; , \quad a_{2-}=L_{-0}^{(0)}
\ee
The charging rate into the eigenstate $\vert s\rangle$ of the DQD from the reservoir $j$ is given by
\be
a_{js} = T_{js}^2 \int_{-\infty}^{+\infty} dt\; {\rm e}^{-i\epsilon_s t} \left\langle {\rm e}^{iH_jt} \, d_{j,0}^{\dagger}\, {\rm e}^{-iH_jt} \, d_{j,0}\right\rangle
\ee
where $j=1,2$, $s=\pm$, and the average is carried out over the equilibrium ensemble (\ref{rho_j}) of the $j^{\rm th}$~reservoir: $\langle\cdot\rangle={\rm tr}\rho_j(\cdot)$.  With Eq.~(\ref{dc}) for $l=0$ and Eq.~(\ref{c_t}), we find
\bea
&&\left\langle {\rm e}^{iH_jt} \, d_{j,0}^{\dagger}\, {\rm e}^{-iH_jt} \, d_{j,0}\right\rangle
 = \int_0^{\pi} dk \, \phi_k(0) \; {\rm e}^{i\epsilon_kt} \nonumber\\
&& \qquad\qquad\times \int_0^{\pi} dk' \, \phi_{k'}(0)  \left\langle c_{j,k}^{\dagger}\, c_{j,k'}\right\rangle
\eea
Using the average (\ref{<c+c>}), we obtain
\be
a_{js} = 2\pi \, T_{js}^2 \int_0^{\pi} dk \, \phi_k(0)^2 \, \delta(\epsilon_k-\epsilon_s) \, f_j(\epsilon_k)
\ee
hence Eq.~(\ref{ajs}) with
\be
\Gamma_{js} = 2\pi \, T_{js}^2 \int_0^{\pi} dk \, \phi_k(0)^2 \, \delta(\epsilon_k-\epsilon_s) 
\label{Gjs}
\ee
which is proportional to the local density of states at the edge $l=0$ of the reservoir in contact with the DQD and at the energy $\epsilon_s$ of the charging transition $s'=0\to s=\pm$.  With the expression (\ref{phi_k}) of the eigenfunction at $l=0$ and the corresponding energy eigenvalue (\ref{E_k}), the rate (\ref{Gjs}) becomes
\be
\Gamma_{js} = T_{js}^2 \, \frac{2}{\gamma} \, \sqrt{1-\left(\frac{\epsilon_s}{2\gamma}\right)^2}
\label{Gjs_bis}
\ee
In the wide-band approximation for which $\vert\epsilon_s\vert\ll2\gamma$, the local density of states is evaluated by Eq.~(\ref{LDOS}) in the middle of the band and we get the result (\ref{Gjs_wb}).

On the other hand, the discharging rates (\ref{bjs}) determine the following elements of the matrix (\ref{L}):
\be
b_{1+}=L_{0+}^{(+)} \; , \quad b_{1-}=L_{0-}^{(+)} \; , \quad b_{2+}=L_{0+}^{(0)} \; , \quad b_{2-}=L_{0-}^{(0)}
\ee
The discharging rate into the eigenstate $\vert s\rangle$ of the DQD from the reservoir $j$ is given by
\be
b_{js} = T_{js}^2 \int_{-\infty}^{+\infty} dt\; {\rm e}^{i\epsilon_s t} \left\langle {\rm e}^{iH_jt} \, d_{j,0}\, {\rm e}^{-iH_jt} \, d_{j,0}^{\dagger}\right\rangle
\ee
with $j=1,2$ and $s=\pm$.  The calculation is similar as in the previous one, using instead the average (\ref{<cc+>}) to get the discharging rate (\ref{bjs}) with Eq.~(\ref{Gjs_wb}) in the wide-band approximation.  

This ends the calculation of the transition rates due to the perturbations of the tunneling interactions $V_{1{\rm A}}$ and $V_{2{\rm B}}$ between the DQD and its reservoirs.  There remains to calculate the rates due to the capacitive coupling with the QPC, which is done in Appendix~\ref{AppE}.

\section{Calculation of the nonequilibrium correlation functions}
\label{AppE}

The capacitive coupling of the DQD with the QPC is again treated perturbatively at second order and in the rotating-wave approximation, but the QPC is supposed to be in the nonequilibrium steady state (\ref{rho_C}).  At the Hamiltonian level of description, the capacitive coupling is expressed with the interaction operator (\ref{VABC}), which has the form
\be
V_{\rm ABC} = S \, R
\ee
\begin{widetext}
\noindent with the subsystem operator $S=U_{\rm A} \, d_{\rm A}^{\dagger} d_{\rm A} + U_{\rm B} \, d_{\rm B}^{\dagger} d_{\rm B}$ and the QPC operator $R=d_{3,0}^{\dagger} d_{4,0} +  d_{4,0}^{\dagger} d_{3,0}$.  The transition rates associated with this interaction are given by
\be
c_{ss'} =L_{ss'}^{(0)}= \tilde C(\omega_{ss'}) \vert\langle s\vert S\vert s'\rangle\vert^2
\label{css}
\ee
with $s=-s'=\pm$, $\omega_{ss'}=\epsilon_s-\epsilon_{s'}$,
\be
\vert\langle +\vert S\vert -\rangle\vert^2 =\vert\langle -\vert S\vert +\rangle\vert^2 = U_{+-}^2 = U_{-+}^2 = \frac{1}{4}\,(U_{\rm A}-U_{\rm B})^2 \sin^2\theta
\label{S+-}
\ee
and the spectral function
\be
\tilde C(\omega) = \int_{-\infty}^{+\infty} dt \, {\rm e}^{-i\omega t} \langle \tilde R(t) \, \tilde R\rangle
\ee
where $\tilde R = R-\langle R \rangle$, $\langle\cdot\rangle={\rm tr}\rho_{\rm C}(\cdot)$, and
\be
X(t) \equiv {\rm e}^{iH_{\rm C} t} X\,  {\rm e}^{-iH_{\rm C} t}
\ee
Using the expression of the operator $R$ and Wick's lemma, the spectral function becomes
\bea
\tilde C(\omega) &=& \int_{-\infty}^{+\infty} dt \, {\rm e}^{-i\omega t} \left[ \langle d_{3,0}^{\dagger}(t) \, d_{4,0}\rangle \langle d_{4,0}(t) \, d_{3,0}^{\dagger}\rangle + \langle d_{3,0}^{\dagger}(t) \, d_{3,0}\rangle \langle d_{4,0}(t) \, d_{4,0}^{\dagger}\rangle \right.\nonumber\\
&&\left. \qquad\qquad\quad\quad + \langle d_{4,0}^{\dagger}(t) \, d_{4,0}\rangle \langle d_{3,0}(t) \, d_{3,0}^{\dagger}\rangle+ \langle d_{4,0}^{\dagger}(t) \, d_{3,0}\rangle \langle d_{3,0}(t) \, d_{4,0}^{\dagger}\rangle\right]
\eea
The correlation functions of the creation-annihilation operators are obtained as
\bea
&&\langle d_{3,0}^{\dagger}(t) \, d_{4,0}\rangle = -\frac{1}{2\pi} \int_0^{\pi} dk \, \vert t_k\vert^2 \, {\rm e}^{i\epsilon_kt} \frac{\gamma}{T_{\rm C}} \left( {\rm e}^{ik} f_{3k} + {\rm e}^{-ik} f_{4k}\right)\\
&&\langle d_{4,0}(t) \, d_{3,0}^{\dagger}\rangle = -\frac{1}{2\pi} \int_0^{\pi} dk \, \vert t_k\vert^2 \, {\rm e}^{-i\epsilon_kt} \frac{\gamma}{T_{\rm C}} \left[ {\rm e}^{ik} (1-f_{3k}) + {\rm e}^{-ik} (1-f_{4k})\right]\\
&&\langle d_{3,0}^{\dagger}(t) \, d_{3,0}\rangle = \frac{1}{2\pi} \int_0^{\pi} dk \, \vert t_k\vert^2 \, {\rm e}^{i\epsilon_kt}  \left( \frac{\gamma^2}{T_{\rm C}^2}\, f_{3k} + f_{4k}\right)\\
&&\langle d_{4,0}(t) \, d_{4,0}^{\dagger}\rangle = \frac{1}{2\pi} \int_0^{\pi} dk \, \vert t_k\vert^2 \, {\rm e}^{-i\epsilon_kt} \left[ 1-f_{3k} + \frac{\gamma^2}{T_{\rm C}^2}\, (1-f_{4k})\right]
\eea
and similar expressions with transposed indices $3$ and $4$.  As a consequence, we have that
\bea
\tilde C(\omega) &=& \frac{1}{2\pi}\int_0^{\pi}dk\int_0^{\pi}dq \, \vert t_k\vert^2 \vert t_q\vert^2 \, \delta(\epsilon_k-\epsilon_q-\omega) \left\{ \frac{2\,\gamma^2}{T_{\rm C}^2}\left[\cos(k+q)+1\right] \left[f_{3k}\, (1-f_{3q})+f_{4k}\, (1-f_{4q})\right] \right.\nonumber\\
&&\left.\qquad\qquad\qquad\qquad\qquad\qquad\qquad+\left[\frac{\gamma^4}{T_{\rm C}^4} +\frac{2\,\gamma^2}{T_{\rm C}^2}\cos(k-q)+1\right] \left[f_{3k}\, (1-f_{4q})+f_{4k}\, (1-f_{3q})\right] \right\}
\eea

In the wide-band approximation, the transmission coefficients as well as the functions $\cos(k\pm q)$  should be evaluated at the values of the wavenumbers $0\leq k,q\leq\pi$ corresponding to the middle of the energy band.  Given the dispersion relation (\ref{E_q}), the only possibility is $k=q=\pi/2$.  Therefore, $\cos(k+q)=-1$, so that the first term is negligible in the wide-band approximation.
On the other hand, $\cos(k-q)=1$, and, using Eq.~(\ref{T_0}), we find
\be
\tilde C(\omega) \simeq \frac{1}{2\pi}\, \frac{4\, \gamma^2}{(T_{\rm C}^2+\gamma^2)^2}  \int_{-\infty}^{+\infty} d\epsilon \, \left\{ f_{3}(\epsilon)\left[1-f_{4}(\epsilon-\omega)\right]+f_{4}(\epsilon)\left[1-f_{3}(\epsilon-\omega)\right]\right\} 
\ee
\end{widetext}
The integral of the first term is evaluated as follows:
\be
\int_{-\infty}^{+\infty} d\epsilon \, f_{3}(\epsilon)\left[1-f_{4}(\epsilon-\omega)\right] = \frac{\omega-\Delta\mu_{\rm C}}{{\rm e}^{\beta(\omega-\Delta\mu_{\rm C})}-1}
\ee
with $\Delta\mu_{\rm C}=\mu_3-\mu_4$ and the other similarly.  Using the local density of states in the middle of the band given by Eq.~(\ref{LDOS}) and the dimensionless contact transparency (\ref{kappa}), we finally get
\be
\tilde C(\omega) \simeq \frac{8\pi\, g^2}{(1+\kappa)^2}\, \left[\frac{\omega-\Delta\mu_{\rm C}}{{\rm e}^{\beta(\omega-\Delta\mu_{\rm C})}-1}+\frac{\omega+\Delta\mu_{\rm C}}{{\rm e}^{\beta(\omega+\Delta\mu_{\rm C})}-1}\right]
\ee
Combining with Eq.~(\ref{S+-}), the expressions (\ref{c+-}) are thus obtained for the transition rates (\ref{css}).

\section{Inequalities deduced from the fluctuation theorems}
\label{AppF}

Here, the inequalities (\ref{inequal}) and (\ref{inequal2}) are proved using Jensen's inequality according to which
\be
\langle f(X)\rangle \geq f(\langle X\rangle)
\ee
for any convex function $f(X)$ and any statistical average $\langle\cdot\rangle$ over the probability distribution of the random variables $X$.\cite{CT06}  The convex function is here taken as $f(X)=\exp X$.

For $X=-\tilde A \, n$ and the statistical average $\langle\cdot\rangle_t=\sum_{n} p(n,t)(\cdot)$ over the probability distribution of the number $n$ of electrons transferred in the DQD, we find
\be
\langle {\rm e}^{-\tilde A \, n}\rangle_t  \geq  {\rm e}^{-\tilde A \langle n\rangle_t}
\ee
By the univariate fluctuation theorem (\ref{FT1}), we have that
\bea
\langle {\rm e}^{-\tilde A \, n}\rangle_t  &=& \sum_n p(n,t) \, {\rm e}^{-\tilde A \, n} \nonumber\\
&\simeq& \sum_n p(-n,t) = \sum_n p(n,t)=1
\eea
hence the inequality (\ref{inequal}).

The other inequality (\ref{inequal2}) results from the bivariate fluctuation theorem of fundamental origin
\be
\frac{p(n,n_{\rm C},t)}{p(-n,-n_{\rm C},t)}\simeq {\rm e}^{A_{\rm D} n+A_{\rm C}n_{\rm C}} \qquad\mbox{for}\quad t\to\infty
\label{FT2}
\ee
where $n=n_{\rm D}$ is the number of electrons transferred during the time interval $[0,t]$ in the DQD and $n_{\rm C}$ in the QPC, while $A_{\rm D}$ and $A_{\rm C}$ are the basic affinities (\ref{A_D})-(\ref{A_C}) of both circuits.

Since the univariate fluctuation theorem (\ref{FT1}) is here supposed to hold jointly with the bivariate theorem (\ref{FT2}), we get
\begin{widetext}
\be
\sum_{n_{\rm C}} {\rm e}^{-A_{\rm D} n-A_{\rm C}n_{\rm C}} p(n,n_{\rm C},t) \simeq \sum_{n_{\rm C}} p(-n,-n_{\rm C},t)= p(-n,t) \simeq  {\rm e}^{-\tilde A \, n} p(n,t)
\ee
after summing only over $n_{\rm C}$.  Multiplying by $\exp(\tilde A \, n)$ and summing also over $n$, we find
\be
\sum_{n,n_{\rm C}} {\rm e}^{(\tilde A-A_{\rm D}) n-A_{\rm C}n_{\rm C}} p(n,n_{\rm C},t) \simeq \sum_n p(n,t)= 1
\label{eq_F1}
\ee  
\end{widetext}
Jensen's inequality with $X=(\tilde A-A_{\rm D}) n-A_{\rm C}n_{\rm C}$ and the statistical average over the probability distribution $p(n,n_{\rm C},t)$ reads
\be
\langle {\rm e}^{(\tilde A-A_{\rm D}) n-A_{\rm C}n_{\rm C}}\rangle_t  \geq  {\rm e}^{(\tilde A-A_{\rm D})  \langle n\rangle_t-A_{\rm C} \langle n_{\rm C}\rangle_t}
\ee
Since $\langle {\rm e}^{(\tilde A-A_{\rm D}) n-A_{\rm C}n_{\rm C}}\rangle_t \simeq 1$ by Eq.~(\ref{eq_F1}), we obtain the inequality
\be
A_{\rm D}  \langle n\rangle_t + A_{\rm C} \langle n_{\rm C}\rangle_t \geq  \tilde A \langle n\rangle_t
\ee
from which Eq.~(\ref{inequal2}) is deduced after dividing by the time interval $t$ and taking the limit $t\to\infty$. Q.~E.~D.


\end{document}